\documentclass[a4paper]{report}
\usepackage[utf8]{inputenc}
\usepackage[T1]{fontenc}
\usepackage{RJournal}
\usepackage{amsmath,amssymb,array}
\usepackage{booktabs}
\usepackage{multirow}
\usepackage{caption}
\usepackage{subcaption}
\usepackage{comment}

\begin{document}

%% do not edit, for illustration only
\sectionhead{Contributed research article}
\volume{XX}
\volnumber{YY}
\year{20ZZ}
\month{AAAA}

%% replace RJtemplate with your article
\begin{article}
  % !TeX root = RJwrapper.tex

\title{\pkg{lsirm12pl}: An R package for the latent space item response model}
\author{Dongyoung Go, Gwanghee Kim, Jina Park, Junyong Park, Minjeong Jeon, Ick Hoon Jin}

\maketitle

\abstract{
% An abstract of less than 150 words.
% \textcolor{blue}{
% The latent space item response model (LSIRM; \citet{jeon2021mapping}) extends the Rasch model \citep{rasch1961} by introducing $D$-dimensional latent Euclidean space, $\mathbb{R}^{D}$. This model is useful for showing interactions between respondents and items in item response data by embedding both items and respondents in a shared, unobserved metric space. LSIRM is valuable for understanding complex respondent-item interactions and improving interpretability. The R package \CRANpkg{lsirm12pl} implements Bayesian estimation of LSIRM and its extensions for different response types, model specifications, and missing data. It offers methods to enhance model utilization, such as clustering item positions in an estimated interaction map. Additionally, \CRANpkg{lsirm12pl} provides summary and plotting options to assess and process estimated results. We provide an overview of the methodological basis of LSIRM and describe its extensions in the package. We demonstrate the practical application of \CRANpkg{lsirm12pl} using real data examples included in the package.
% }

The item response model in latent space (LSIRM; \citet{jeon2021mapping}) uncovers unobserved interactions between respondents and items in the item response data by embedding both in a shared latent metric space. The R package \CRANpkg{lsirm12pl} implements Bayesian estimation of the LSIRM and its extensions for various response types, base model specifications, and missing data handling. Furthermore, \CRANpkg{lsirm12pl} package provides methods to improve model utilization and interpretation, such as clustering item positions on an estimated interaction map. The package also offers convenient summary and plotting options to evaluate and process the estimated results. In this paper, we provide an overview of the LSIRM's methodological foundation and describe several extensions included in the package. We then demonstrate the use of the package with real data examples contained within it.
}

% - for all figures and tables, add the model name (e.g., 1pl LSIRM) and data name. And add more explanations if possible. 
% Right now, the captions are often too brief, and it is not clear which models the results are for. 

% - for 2pl model and continuous model illustrations, explanations are very brief. I think there should be more explanations 

% - related to the above, I also asked you to add some explanations of the tools/methods we are using. We should not assume readers already know them. 
% The reviewer’s comment that the paper looks like a vignette rather than a paper is related to this issue. 

% - how did you handle the 2PL model for identifiability?
% Typically, one of the alphas should be fixed to 1, or the variance of the latent trait should be fixed to 1.

\section{Introduction}
Item response theory (IRT) models are a widely-used statistical approach to analyze assessment data in various fields for e.g., medical, educational, psychological, health, and marketing research \citep{An2014, Zanon2016}.  IRT models are designed to establish a relationship between observed item response data and unobserved person characteristics, commonly referred to as latent traits, e.g., competencies, attitudes, or personality \citep{Ayala2009, Brzezinska2018}. IRT models can predict the probability of a correct (or positive) response as a function of the respondents' latent traits and item features, such as item difficulty and discrimination. Additional technical details of IRT models are provided in the subsequent section.
%The 2PL model includes an additional set of item parameters, called item discrimination parameters, which measure test items' ability to discriminate between respondents with similar ability levels. 

Several R packages are available for estimating IRT models. The \CRANpkg{ltm} package  \citep{Rizopoulos2006} is available to analyze dichotomous and polytomous item response data, including a oneparameter logistic (1PL) model (or the Rasch model), a two-parameter logistic (2PL) model, a three-parameter model \citep{rasch:60, birnbaum1968} and a graded response model \citep{Samejima1968}. The \CRANpkg{eRm} package \citep{Mair2007} estimates various extensions of the Rasch model, such as the rating scale model (RSM) \citep{Andrich1978}, partial credit model (PCM) \citep{masters1982}, linear logistic test model (LLTM) \citep{scheiblechner1972}, the linear rating scale model (LRSM) \citep{fischer1991}, and the linear partial credit model (LPCM) \citep{glas1989, fischer1994}. The \CRANpkg{mirt} package \citep{Chalmers2012} can estimate a wide range of IRT models, including exploratory and confirmatory multidimensional item response models. 
The \CRANpkg{pcIRT} package \citep{Hohensinn2018} provides functions for estimating IRT models for polytomous (nominal) and continuous data, including the multidimensional polytomous Rasch model \citep{Andersen1973} and the continuous rating scale model \citep{Mller1987}, using conditional maximum likelihood (CML) estimation \citep{Baker2023}. 
%The \CRANpkg{brms} package \citep{Burkner2021} offers options to estimate Bayesian IRT models and models for reaction times or rates. %\textcolor{red}{mj: I am not sure about the last package brms. I don't think it is an IRT package, but a more general multilevel modeling package. I would remove it}

Conventional IRT models are typically based on two assumptions: conditional independence and homogeneity. Conditional independence assumes that item responses are independent of each other conditional on respondents' latent traits. The homogeneity assumption is e.g., that respondents with the same ability have the same probability of answering a question correctly and that respondents have the same probability of answering a question with the same item difficulty. However, these assumptions are often violated in practice due to unobserved interactions between respondents and items, e.g., when particular items are more similar to each other (e.g., testlets) or when particular respondents show different probabilities of giving correct responses to certain items compared to other respondents with similar ability (e.g., differential item functioning). Violations of these assumptions can lead to biased parameter estimates and inferences \citep{Chen2007, Braeken2010, Myszkowski2024}. While some existing methods can address known violations prior to data analysis, there is currently no approach that enables the detection or management of unknown sources of violations in item response analysis, to the best of our knowledge.

\citet{jeon2021mapping} proposed a latent space item response model (LSIRM) that addresses such limitations of conventional IRT models. The LSIRM aims to estimate inherent interactions between respondents and items, alongside the latent traits of both. A key feature is its visual representation of these interactions in a low-dimensional latent space, called an \textit{interaction map} in the form of distances between them. Interaction maps offer a clear and intuitive interpretation of complex respondent-item relationships, allowing users to identify patterns and clusters based on spatial proximity on the map. Additional details of the LSIRM are provided in a later sections.
% and methods by introducing an interaction map, which embeds both items and respondents in a shared and unobserved metric space. 
%LSIRM is a powerful tool for analyzing item response data, which usually consists of the respondents' answers to questions or statements on tests, surveys, or questionnaires. 
% The probability of a correct response is modeled as a function of the distance between the respondent and the item's latent positions in the latent space, given the respondent's ability and the item's difficulty. %In addition to addressing the assumptions of conditional independence and homogeneity, 
% The estimated interaction map also supplies useful insights and diagnostic information on both items and respondents. 

This paper presents the \CRANpkg{lsirm12pl} package in R that offers Bayesian estimation of the LSIRM and its extensions. \citet{jeon2021mapping} focused on the response data for binary items and the Rasch model as the base model, and currently, no package is available to estimate the LSIRMs. To broaden the applicability of latent space item response modeling, \CRANpkg{lsirm12pl}  enables: (1) modeling continuous item responses; (2) missing data handling under different missing mechanisms assumptions \citep{rubin1976inference}; and (3) an extended base model specification using the 2PL model both for binary and continuous item response data. The package also offers options to cluster latent positions of items in the estimated interaction map using spectral clustering and the Neyman-Scott process model. The \CRANpkg{lsirm12p} supplies convenient summary and plotting options for interaction maps, model assessment, diagnosis, and result process and interpretation, aiming to improve the utilization of the LSIRM in practice. 
%A function to visualize unobserved interactions between items and respondents in a two-dimensional latent space, 
% Details of the clustering methods are described in Section \ref{lsirm_1pl_data}. 

The subsequent sections of the paper are structured as follows. To begin, we  provide a concise overview of the 1PL and 2PL IRT models. We then delve into the LSIRM for binary response data and demonstrate how to fit the model with the package functions using a real dataset. Next, we extend the LSIRM to accommodate continuous data and provide guidance on fitting this extended model using the \CRANpkg{lsirm12pl} package. In the end, we conclude the paper with final remarks and a discussion of future developments.

%\textcolor{red}{[[mj: clustering part seems missing/ JA: There is no separate section for clustering and is included in the illustrated example section. also description of clustering part included in previous paragraph]]}

\section{Item Response Theory Models}\label{irt}

IRT models are essential for analyzing item response data, which comprises respondents' answers to items on tests, surveys, or questionnaires. This section briefly discuss two widely utilized IRT models for dichotomous item response data.

\subsection{1PL IRT Model}

The 1PL model, also known as the Rasch model \citep{rasch1961}, is a classic IRT model for analyzing dichotomous item response data. Suppose $\boldsymbol{Y} = \left\{ y_{k,i} \right\} \in \left\{ 0, 1 \right\}^{N \times P}$ is the $N \times P$ binary item response matrix under analysis, where $y_{k,i}=1$ indicates a correct (or positive) response of the respondent $k$ to the item $i$. In the Rasch model, the probability of the correct response for item $i$ given by respondent $k$ is given as follows:
\begin{equation*}
\text{logit}(\mathbb{P}(y_{k,i}=1|\theta_{k},\beta_{i}))=\theta_{k} +\beta_{i}, \; \; \theta_{k} \sim N(0, \sigma^2), 
\label{rasch_formula}
\end{equation*}
where $\theta_k \in \mathbb{R}$ is the respondent intercept parameter of respondent $k$, $\beta_i \in \mathbb{R} $ is the item intercept parameter of item $i$. The person intercept parameter represents the latent trait of interest, such as cognitive ability. The item intercept parameter represents the item easiness (or minus difficulty). As the respondent's ability increases, her/his likelihood of giving correct responses increases. On the other hand, the likelihood of correct responses decreases as the difficulty of the item increases. Note that based on the model, the probability of a respondent's giving a correct response to an item is a function of the two parameters -- the person's ability and the item's difficulty. This means that respondents with the same level of ability are assumed to have the same probability of giving a correct response to an item. Similarly, items with the same level of difficulty are assumed to have the same probability of being correctly responded. In other words, interactions between persons and items are not allowed in this model. However, in reality, respondents with the same level of ability and items with the same level of difficulty may show a different success probability due to unobserved characteristics they may have, violating the assumptions. %, e.g., mentoring, training for respondents and diff

\subsection{2PL IRT Model}

The 2PL model \citep{birnbaum1968} extends the 1PL model by incorporating a discrimination parameter for each item. This parameter reflects the ability of the item to differentiate among respondents with similar abilities. The item discrimination parameters are also considered as item slopes, indicating how quickly the probability of correct responses increases as the respondent's abilities increase \citep{An2014}. A larger discrimination parameter indicates that the item is better at distinguishing between respondents with similar ability levels. In the 2PL model, the probability of person $k$ giving a correct response to item $i$ is given as follows:
\begin{equation}
\text{logit}(\mathbb{P}(y_{k,i}=1|\theta_{k}, \alpha_{i}, \beta_{i}))=\alpha_{i}\theta_{k} + \beta_{i}, \; \; \theta_{k} \sim N(0, \sigma^2), 
\label{birnbaum2pl_formula}
\end{equation}
where $\alpha_i \in \mathbb{R}$ is the discrimination parameter for the item $i$. For model identifiability, one of the item slope parameters is fixed at 1, e.g., $\alpha_{1} =1$. With the term $\alpha_{i}\theta_{k}$, the 2PL model allows for an interaction between respondents and items to some degree. However, it is not likely to capture all interactions between respondents and items that might not depend on respondents' ability \citep{jeon2021mapping}. 

\section{Standard LSIRM: 1PL LSIRM For Dichotomous Data}\label{1pl}

The LSIRM \citep{jeon2021mapping} has been proposed as an extension of the conventional IRT models. 
The key idea of the LSIRM is to place respondents and items in a low-dimensional, shared metric space, called an interaction map, so that their unobserved interactions can be captured in the form of distances between them. Below we provide a brief overview of the LSIRM in its original form, presented for dichotomized data. 

\subsection{Statistical framework}

% Consider the $N \times P$ binary item response matrix $\boldsymbol{Y} = \left\{ y_{k,i} \right\} \in \left\{ 0, 1 \right\}^{N \times P}$, where $y_{k,i}$ is a response of the respondent $k$ to the item $i$. Let $y_{k,i} = 1$ indicate a correct (or positive) response of the respondent $k$ to item $i$, while $y_{k,i}=0$ indicates an incorrect (or negative) response. 
To capture unobserved interactions between respondents and items, the original LSIRM \citep{jeon2021mapping} embeds items and respondents in an interaction map, i.e., a $D$-dimensional latent Euclidean space. Note that the interaction map is used as a tool to represent pairwise interactions between respondents and items; thus, the dimensions of an interaction map do not represent any specific quantity or have a substantive meaning. The original LSIRM is built on the 1PL IRT model; thus, we refer to the original LSIRM as the 1PL LSIRM.
% and to alleviate the conditional independence and homogeneity assumptions of conventional IRT models, 
%extends the Rasch model \citep{rasch1961} by 
%introduces $D$-dimensional latent Euclidean space $\mathbb{R}^{D}$ into the Rasch model. 
%a shared and unobserved $D$-dimensional latent Euclidean space, called an interaction map. 
%The interaction map captures and represents pairwise interactions between respondents and items. 
% with item intercept parameters;  %only to characterize item characteristics

The 1PL LSIRM assumes that the probability of giving a correct answer to item $i$ by the respondent $k$ is determined by a linear combination of the main effect of the respondent $k$, the main effect of the item $i$, and the pairwise distance between the latent position of item $i$ and the latent position of respondent $k$. Then, the model is given by: 
\begin{equation}
\text{logit}(\mathbb{P}(y_{k,i}=1|\theta_{k},\beta_{i},\gamma,\boldsymbol{z_{k},w_{i}}))=\theta_{k}+\beta_{i}-\gamma d (\boldsymbol{z_{k}},\boldsymbol{w_{i}}),
\label{1pl_formula}
\end{equation}
where $\theta_k \in \mathbb{R}$ and $\beta_i \in \mathbb{R}$ represent the respondent's latent trait and the item's easiness, respectively, same as in the conventional 1PL model. These two terms can also be seen as the main effects of respondents and items. The third term, $-\gamma d (\boldsymbol{z_{k}},\boldsymbol{w_{i}})$, captures the interactions between respondents and items, where $\boldsymbol{z_{k}}\in\mathbb{R}^{D}$ and $\boldsymbol{w_{i}}\in\mathbb{R}^{D}$ are the latent position of the respondent $k$, and the latent position of the item $i$, respectively, where  $\gamma \geq 0$ is the weight of the distance term $d(z_{k},w_{i})$. For a distance function $d:$ $\mathbb{R}^{D} \times \mathbb{R}^{D} \mapsto [0,\infty)$, we  use a Euclidean norm (that is, $d(z_{k},w_{i})=||z_k - w_i||$) as a distance function in \code{lsirm1pl} to maintain simplicity and enhance the interpretability of the model \citep{Hoff:2002}. The weight of the distance term $\gamma \geq 0$ indicates the amount of interactions between respondents and items in the data, such that large $\gamma$ implies stronger evidence for respondent by item interactions in the data. In contrast, near zero $\gamma$ implies little evidence of respondent-item interactions in the data, thus suggesting that one may go with the conventional IRT model without the interaction term. 
%$\theta_k \in \mathbb{R}$ is the respondent intercept parameter of respondent $k$, $\beta_i \in \mathbb{R} $ is the item intercept parameter of item $i$, 
% \textcolor{blue}{In summary, the one of key features of a LSIRM is an estimate of the main effects, $\beta_i$ and $\theta_k$, along with an interaction map. $\theta_k$ represents the ability of the respondent $k$, with larger values indicating a higher probability of getting the correct answer. $\beta_i$ represents the difficulty of the item $i$, with larger values indicating that more respondents are more likely to answer correctly. LSIRM also improves accessibility and understanding of unobserved interactions between items and respondents with interaction maps that visualize these interactions within a two-dimensional latent space.}

The likelihood function of the observed data with the 1PL LSIRM is 
\begin{equation*}
\mathbb{L}\left(\boldsymbol{Y=y | \theta, \beta}, \gamma, \boldsymbol{Z, W} \right) = \prod_{K=1}^{N} \prod_{i=1}^{P} \mathbb{P}(Y_{k,i} = y_{k,i} | \theta_{k}, \beta_{i}, \gamma, \boldsymbol{z_{k}, w_{i}}),
\label{likelihood}
\end{equation*}
where $\boldsymbol{\theta} = \left(\theta_{1}, \ldots, \theta_{N}\right)$, $\boldsymbol{\beta}=\left(\beta_{1}, \ldots, \beta_{P}\right)$, $\boldsymbol{Z} = \left(\boldsymbol{z_{1}}, \ldots, \boldsymbol{z_{N}}\right)$ and $\boldsymbol{W}=\left(\boldsymbol{w_{1}}, \ldots, \boldsymbol{w_{P}}\right)$. Here, item responses are assumed to be independent conditional on the positions of respondents and items in an interaction map, as well as the main effects of respondent and item. This means that the traditional conditional independence assumption is alleviated with the model by accounting for respondent-item interactions. The detailed parameter estimations are described in the following Section.
% 1PL posteior
% &= \mathbb{L}\left(\boldsymbol{Y=y | \theta, \beta}, \gamma, \boldsymbol{Z, W} \right) \pi(\boldsymbol{\theta})\pi(\boldsymbol{\beta})\pi(\boldsymbol{\gamma})\pi(\boldsymbol{Z})\pi(\boldsymbol{W}) \\

\subsection{Parameter Estimation} \label{lsirm_1pl_est}

%All model parameter estimates  are  based on posterior samples. We set a normal and a multivariate normal distribution (MVN) with mean 0 as priors for $\theta_{k}, \beta_{i}, \boldsymbol{z_k}$ and $\boldsymbol{w_i}$. Conjugate inverse-Gamma distributions are used as priors for the variance parameters of the random effect $\theta_{k}$, $k=1,..,N$. We use a log-normal distribution for the prior of the weight parameter $\gamma$ due to its non-negative property. 
The R package \CRANpkg{lsirm12pl} applies a fully Bayesian approach using the Markov chain Monte Carlo (MCMC) for estimation of the LSIRM. Following is the posterior distribution of the 1PL LSIRM.
\begin{align*}
\pi\left(\boldsymbol{\theta, \beta}, \gamma, \boldsymbol{Z, W} | \boldsymbol{Y=y} \right) & \propto \prod_{k=1}^{N} \prod_{i=1}^{P} \mathbb{P}(Y_{k,i} = y_{k,i} | \theta_{k}, \beta_{i}, \gamma, \boldsymbol{z_{k}, w_{i}}) \\
&\times \prod_{k=1}^N \pi(\theta_k) \times \prod_{i=1}^P \pi(\beta_i) \times \pi(\gamma) \times \prod_{k=1}^N \pi(\boldsymbol{z_{k}}) \prod_{i=1}^P \pi(\boldsymbol{w_{i}})
\label{posterior}
\end{align*}
The priors for the model parameters are specified as follows:
\begin{equation*}\label{1pl-prior}
\begin{split}
\theta_{k} \mid \sigma^{2} & \sim N(0,\sigma^{2}),\ \sigma^{2}>0\\
\beta_{i} \mid \tau_{\beta}^{2} & \sim N(0,\tau_{\beta}^{2}),\ \tau_{\beta}^{2}>0\\
\log \gamma \mid \mu_{\gamma},\tau_{\gamma}^2 & \sim N(\mu_{\gamma},\tau_{\gamma}^2),\ \mu_{\gamma}\in\mathbb{R},\ \tau_{\gamma}^2>0\\
\sigma^{2} \mid a_{\sigma},b_{\sigma} & \sim\text{Inv-Gamma}(a_{\sigma},b_{\sigma}),a_{\sigma}>0,b_{\sigma}>0\\
\boldsymbol{z_{k}} & \sim\text{MVN}_{D}(\boldsymbol{0,I_{D}})\\ 
\boldsymbol{w_{i}} & \sim\text{MVN}_{D}(\boldsymbol{0,I_{D}}).
\end{split}
\end{equation*}
where $\boldsymbol{0}$ is a $D$-vector of zeros and $\boldsymbol{I_{D}}$ is $D \times D$ identity matrix. The argument names and default values for the prior specifications in the \CRANpkg{lsirm12pl} are described in our Github site \footnote{https://github.com/jiniuslab/lsirm12pl}. 

To generate posterior samples for $\boldsymbol{\theta,\ \beta,\ \gamma,\ Z}$ and $\boldsymbol{W}$, we use the Metropolis-Hastings-within-Gibbs sampler \citep{chib1995understanding}. The conditional posterior distribution for each parameter is given in Equations in the Github site. We list the arguments, the default values for the jumping rules, and the standard deviations of the Gaussian proposal distributions in the Github site.

The log-odds of the probability of giving correct responses depend on the latent positions through distances, as discussed in \citet{jeon2021mapping}. Because distances are invariant to translations, reflections, and rotations of the positions of respondents and items, the likelihood function is invariant under these transformations. To resolve the identifiability issue of latent positions, the \CRANpkg{lsirm12pl} package applies Procrustes transformation \citep{procruste} as a post-processing of posterior samples, which is a standard procedure in the latent space modeling literature \citep{Hoff:2002, Sewell:2015, jeon2021mapping}.

\subsection{An Illustrated Example}\label{lsirm_1pl_data}

Here, we demonstrate how to apply the 1PL LSIRM to real datasets using the \CRANpkg{lsirm12pl} package. To this end, we use the Inductive Reasoning Developmental Test (TDRI) dataset \citep{TDRI} from the package, which contains item responses from 1,803 Brazilians (52. 5\% female) of ages ranging from 5 to 85 years (M = 15.75; SD = 12.21). TDRI is a pencil-and-paper test consisting of 56 items that are designed to assess developmentally sequenced and hierarchically organized inductive reasoning. 
%\textcolor{red}{mj: we mentioned earlier that we use data included in the package. In that case, the link below is unccessary; just mention that  the data are included in the package. I am sure the data citation is included in the package }
%The dataset is publicly available for reproducibility purposes and can be downloaded from the following link: 
% \begin{center}
% \url{https://figshare.com/articles/dataset/TDRI\_dataset\_csv/3142321}.
% \end{center}

The \CRANpkg{lsirm12pl} package provides several functions for fitting the LSIRM, which requires setting hyperparameter values for prior distributions and/or tuning parameters for MCMC chains. By default, the \code{lsirm} function runs with the default settings unless otherwise specified by the user. The default MCMC run setting includes 15,000 iterations, 2,500 burn-ins, and 5 thinning. The base function for running the LSIRM is 
\begin{center}
\code{lsirm(A $\sim$ <term 1>(<term 2>, <term 3>, ...))}    
\end{center}
%, and prints the parameter values every 500 iterations \textcolor{red}{mj: posterior samples?}. 
where \code{A} is an item response matrix to be analyzed, \code{<term1>} is for the model option -- either `lsirm1pl' or `lsirm2pl', while \code{<term 2>} and \code{<term 3>} are other specific options for the chosen model, which are detailed in the documentation of the \CRANpkg{lsirm12pl} package. The following is an example of how to fit the 1PL LSIRM to the TDRI dataset (with no missing) with a default estimation setting with 4 MCMC chains using 2 multi-core processors. Additional details of other default values can be found in the documentation of the \CRANpkg{lsirm12pl} package.
%\textcolor{blue}{The model was fitted with 4 MCMC chains using 2 multi-core processors, and all other options were set to their default values.} 

\begin{example}
R > library("lsirm12pl")
R > data <- lsirm12pl::TDRI
R > data <- data[complete.cases(data),]
R > head(data)
  i1 i2 i3 i4 ... i54 i55 i56
1  1  1  1  1  ... 0   0   0
2  1  1  1  1  ... 0   0   0
3  1  1  1  1  ... 0   0   0
4  1  1  1  1  ... 0   0   0
5  1  1  1  1  ... 0   0   0

R > lsirm_result <- lsirm(data ~ lsirm1pl(chains = 4, multicore = 2, seed = 2025))
\end{example}
The estimation results of the model parameters $\boldsymbol{\theta,\ \beta},\ \gamma,\ \boldsymbol {Z}$, and $\boldsymbol{W}$ are stored in individual lists per chain, while all results across the chains are stored in a single list. %The summary results are provided with the \code{summary()} function. 

The \code{summary()} function generates a summary of the first chain by default, but users can obtain summaries of other chains by setting the \code{chain.idx} option. The function provides posterior estimates of the ``covariate coefficients'' (model parameters such as $\beta$), allowing the users to select the \code{"mean"}, \code{"median"}, or \code{"mode"} through the \code{estimate} option. It also provides the highest posterior density interval (HPD) for the model parameters with the \code{CI} option that allows the users to set  different significance levels. Additionally, the function supplies the Bayesian information criterion (BIC) and the maximum log posterior value. When the column names are available in the input data, the \code{summary()} function uses these names in the summary of the results. 
%The summary function prints the Bayesian estimating setting being used, such as MCMC sample size and burn-in period. 
%For instance, in the TDRI dataset, where columns are named i1 through i56, the output summary will use these column names. 
\begin{example}
R > summary(lsirm_result, chain.idx = 1, estimate = 'mean', CI = 0.95)
========================== 
Summary of model 
========================== 

Call:	 lsirm.formula(formula = data ~ lsirm1pl(chains = 4, multicore = 2, seed = 2025)) 
Model:	 lpl LSIRM 
Data type:	 binary 
Variable Selection:	 FALSE 
Missing:	 NA 
MCMC sample of size 15000, after burnin of 2500 iteration 

Covariate coefficients posterior means of chain 1 : 

    Estimate     2.5%   97.5%
i1   6.42354  5.82468  7.0642
...
i56  -1.01604 -2.58822  0.8133

--------------------------- 

Overall BIC (Smaller is better) : 43661.52 

Maximum Log-posterior Iteration:  
      value iter
[1,] -11487 137
\end{example}

The \code{diagnostic()} function checks the convergence of MCMC for each parameter using various diagnostic tools, such as trace plots, posterior density distributions, autocorrelation functions (ACF), and Gelman-Rubin-Brooks plots. The  \code{diagnostic()} function has options:  \code{draw.item}, and \code{gelman.diag}. The \code{draw.item} option in the \code{diagnostic()} function specifies the names and indexes of the parameters to diagnose. The \code{draw.item} option is set to a list where a key represents each parameter such as \code{``beta'', ``theta'', ``gamma'', ``alpha'', ``sigma''}, and \code{``sigma\_sd''}, and the values indicate the indices of these parameters. The indexes can be expressed as vectors. In the following example code, the \code{draw.item} option  is set as \code{list(``beta'' = c(1))} to check the diagnostic of the first index of the beta parameter, i.e. $\beta_1$.  With \code{gelman.diag = TRUE}, the Gelman-Rubin convergence diagnostic, known as potential scale reduction factors (PSRF), is obtained. 
% \textcolor{red}{
% [[mj: it is unclear what "beta" = c(1) means]]
% }

\begin{example}
R > diagnostic(lsirm_result,
           draw.item = list("beta" = c(1)),
           gelman.diag = TRUE)

Potential scale reduction factors:

          Point est. Upper C.I.
beta [i1]       1.01       1.04
\end{example}

\begin{figure}[ht]
    \centering
    \includegraphics[width = .8\textwidth]{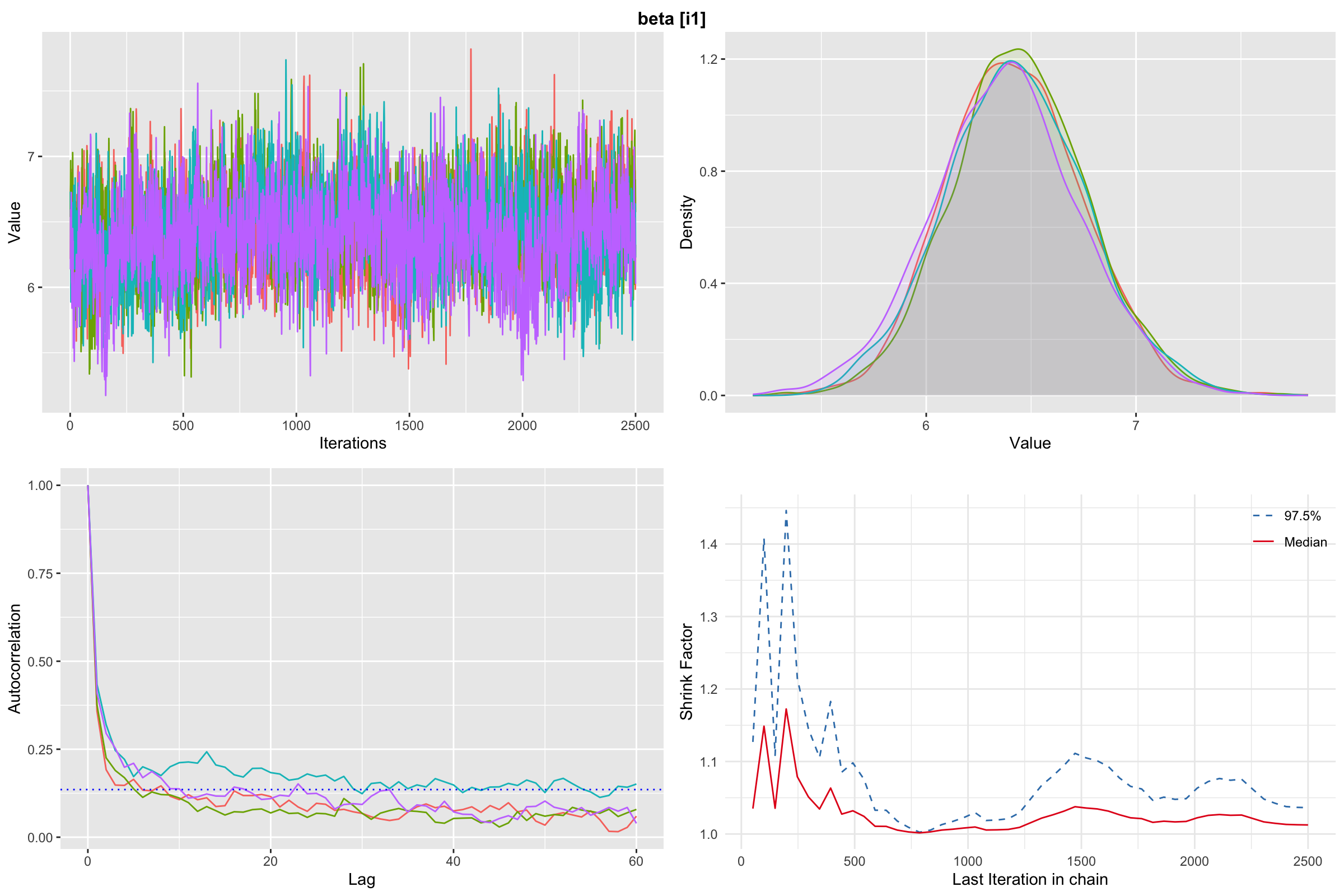}
    \caption{
    Diagnostics of $\beta_1$ using the \code{diagnostic()} function on the results of the 1PL LSIRM fitted to the TDRI dataset. The top left is the trace plot, the top right is the posterior density plot, the bottom left is the autocorrelation plot, and the bottom right is the Gelman-Rubin-Brooks plot. Different colors represent different MCMC chains.
    }
    \label{fig:diagnosis1pl}
\end{figure}

Figure \ref{fig:diagnosis1pl} displays the output of \code{diagnostic()} for $\beta_1$: trace plot (top left), posterior density plot (top right), autocorrelation plot (bottom left), and Gelman-Rubin-Brooks plot (bottom right). Different colors indicate different MCMC chains. Trace plots visualize the mixing of the MCMC chains. In a well-converged model, the trace plots should show that chains fluctuate consistently around a constant value, which indicates the parameter space is thoroughly explored. Density plots display the distribution of the sampled parameter values. In a converged model, the density plots should exhibit smooth, overlapping curves across different chains, which demonstrates that the samples are drawn from the same posterior distribution. Autocorrelation plots measure the correlation between samples at different lags. For a converged model, the autocorrelation should decrease rapidly as the lag increases, which suggests the samples are independent and the parameter space has been effectively explored. Gelman-Rubin-Brooks plots show a shrink factor, known as the potential scale reduction, which compares the variance within each chain to the variance between chains. A shrink factor value close to 1 indicates the within-chain variance is similar to the between-chain variance, providing evidence of good convergence. By examining these diagnostics collectively, the convergence of the model parameters from the LSIRM can be ensured. 
%To diagnose the convergence of the LSIRM, we employ a combination of diagnostic tools, including trace plots, autocorrelation, potential scale reduction (PSR), and density plots. 
%[[mj: the plot says "shrink factor" instead of shrink factor. this should be explained]]
% and ensure the reliability of the statistical inferences derived from the LSIRM.

The \CRANpkg{lsirm12pl} package supplies the \code{gof()} function that assesses the goodness of fit of the LSIRM to the item response data being analyzed. A main diagnostic tool is a boxplot that displays the average posterior `predicted' response value for each item (item-wise means) with the average `observed' response value for each item indicated by red dots. When the model fits well, the red dot is close to the midline of the boxplot. The \code{gof()} function includes a \code{chain.idx} option to choose a specific chain for assessing the goodness of fit. By default, the first MCMC chain is selected (i.e. \code{chain.idx=1}). When the \code{diagnostic()} function shows good convergence for all parameters, it does not matter which chain is chosen because the parameter estimates must be consistent across all chains. For illustration, we use the first MCMC chain by setting \code{chain.idx=1}. % from now on.

% \textcolor{red}{[[mj: but why? it is kind of odd that we use only one chain after fitting the model with multiple chains. do we suggest this practice (using one chain)? if not, please explain what happens if the chain is not chosen (in other words, all chains are used). same for all other examples below. / JA: Multiple chains are only needed for diagnostics. A good convergence with diagnostics is equivalent to all chains showing the same result, so it doesn't matter which chains are used. ]]}

For binary data, the \code{gof()} function additionally offers a receiver operating characteristic (ROC) curve. The ROC curve is a graphical representation that assesses the performance of a binary classifier. The area under the curve (AUC) quantifies the overall ability of the model to distinguish between classes, with values ranging from 0.5 to 1. A higher AUC and a curve close to the top-left corner indicate better performance.
% [[mj: explain a bit about ROC curve. what it is, how to get it, and how to interpret, etc.]]
% 
\begin{example}
R > gof(lsirm_result, chain.idx = 1)
\end{example}
\begin{figure}[ht]
    \centering
    \includegraphics[width=.7\textwidth]{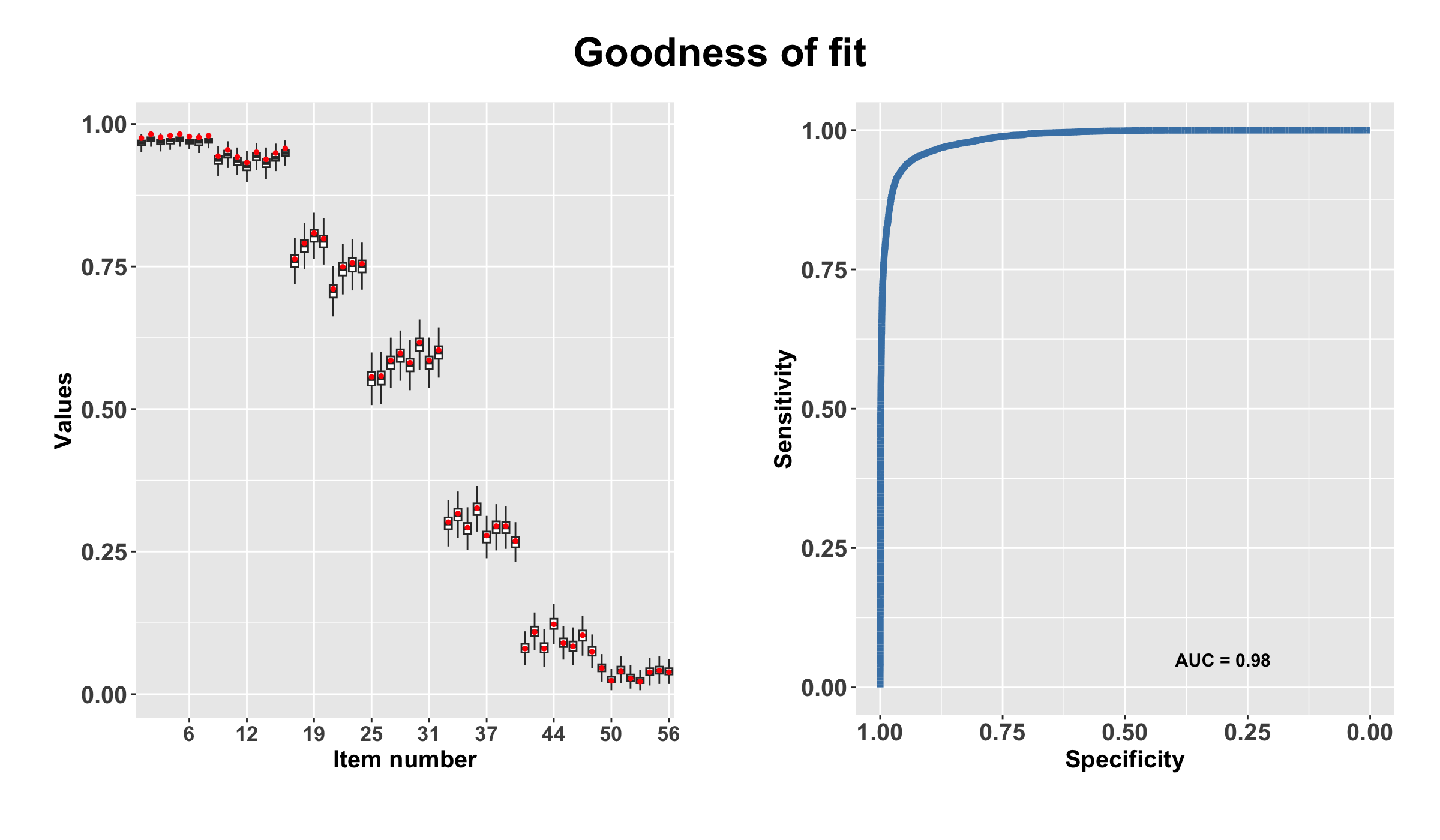}
    \caption{
    Goodness of fit of the 1PL LSIRM for the TDRI data using the \code{gof()} function. The box plots of the average posterior predicted response values for items (item-wise means) with the average observed response value for each item indicated by red dots (left) and the receiver operating characteristic (ROC) curve (right). 
    }
    \label{fig:gof1}
\end{figure}

Figure \ref{fig:gof1} presents the output of the \code{gof()} function for the TDRI dataset, showing the results for the first MCMC chain, including both the boxplot (left) and the ROC curve (right). In this example, all of the red dots are closely located at the midline of the boxplot and the area under the curve (AUC) of the ROC curve approaches 1, suggesting that the fit of the LSIRM to the data is satisfactory. 

% \textcolor{red}{[[mj: actually, the fit of items 18-31 do not look that good. need to mention this / Solve! (Changes the posterior predicted response value calculation method used when drawing a boxplot.)]]}

The main outcome of the LSIRM fitting is the estimates of the respondent and item main effects and the interaction map. For a visual summary of these outputs, the user can use the \code{plot()} function with the \code{option} argument. % for two main purposes: (1) to summarise parameter estimates for $\beta_i$ and $\theta_k$, and (2) to illustrate interaction maps. To summarise the ability of respondents and the difficulty of items, the \code{option} argument can be set to \code{"beta"} and \code{"theta"}, respectively. Note that, as discussed in the LSIRM overview, respondents with high $\theta_k$ are more likely to answer the item correctly, while items with high $\beta_i$ are more likely to be answered correctly by most individuals.

\begin{example}
R > plot(lsirm_result, option = "beta", chain.idx = 1)
R > plot(lsirm_result, option = "theta", chain.idx = 1)
\end{example}

\begin{figure}[ht]
\centering
    \begin{subfigure}[b]{0.3\textwidth}
        \centering
        \includegraphics[width=\textwidth]{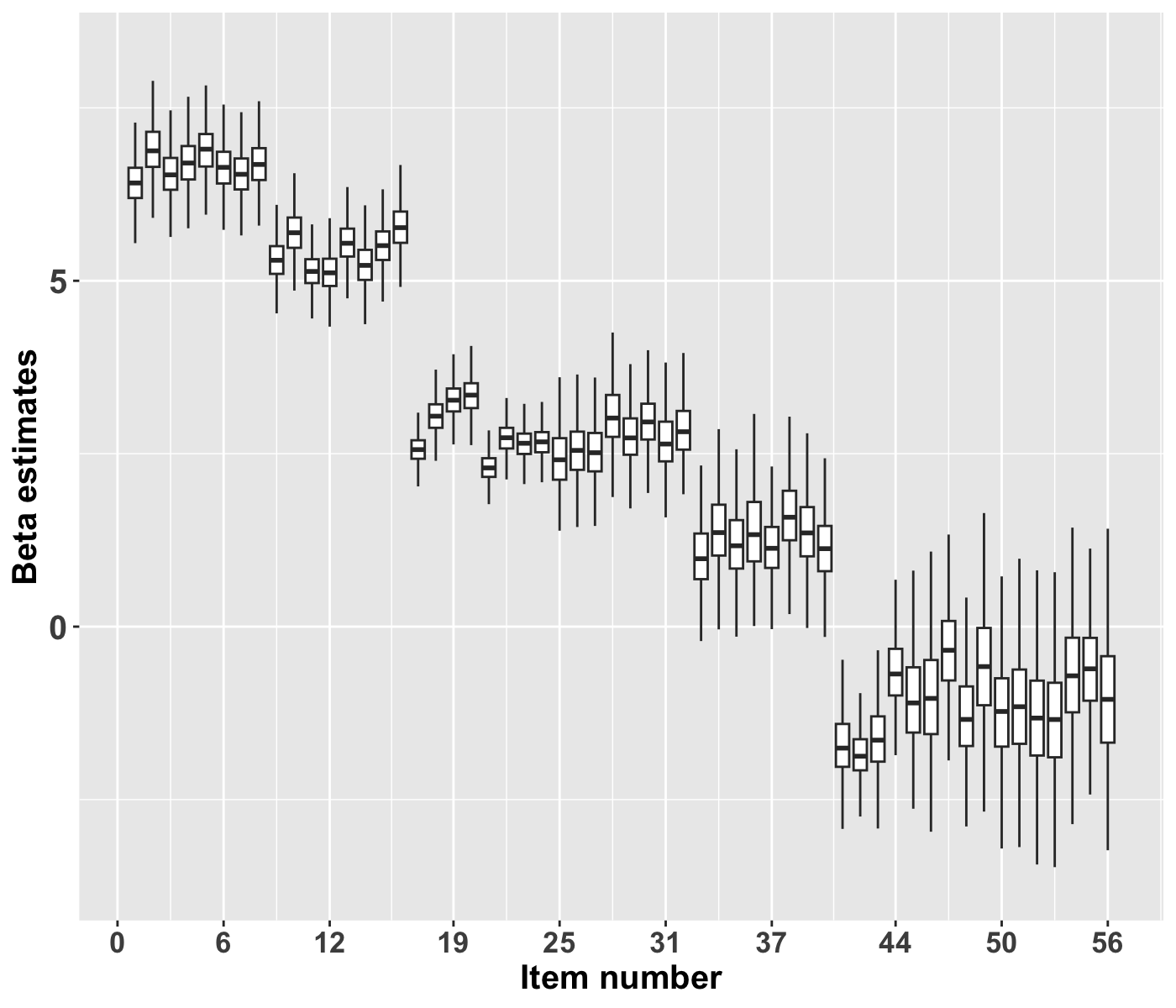}
        \caption{$\beta_i$} \label{fig:latent_parama}
    \end{subfigure}
    \hspace{0.05\textwidth}
    \begin{subfigure}[b]{0.3\textwidth}
        \centering
        \includegraphics[width=\textwidth]{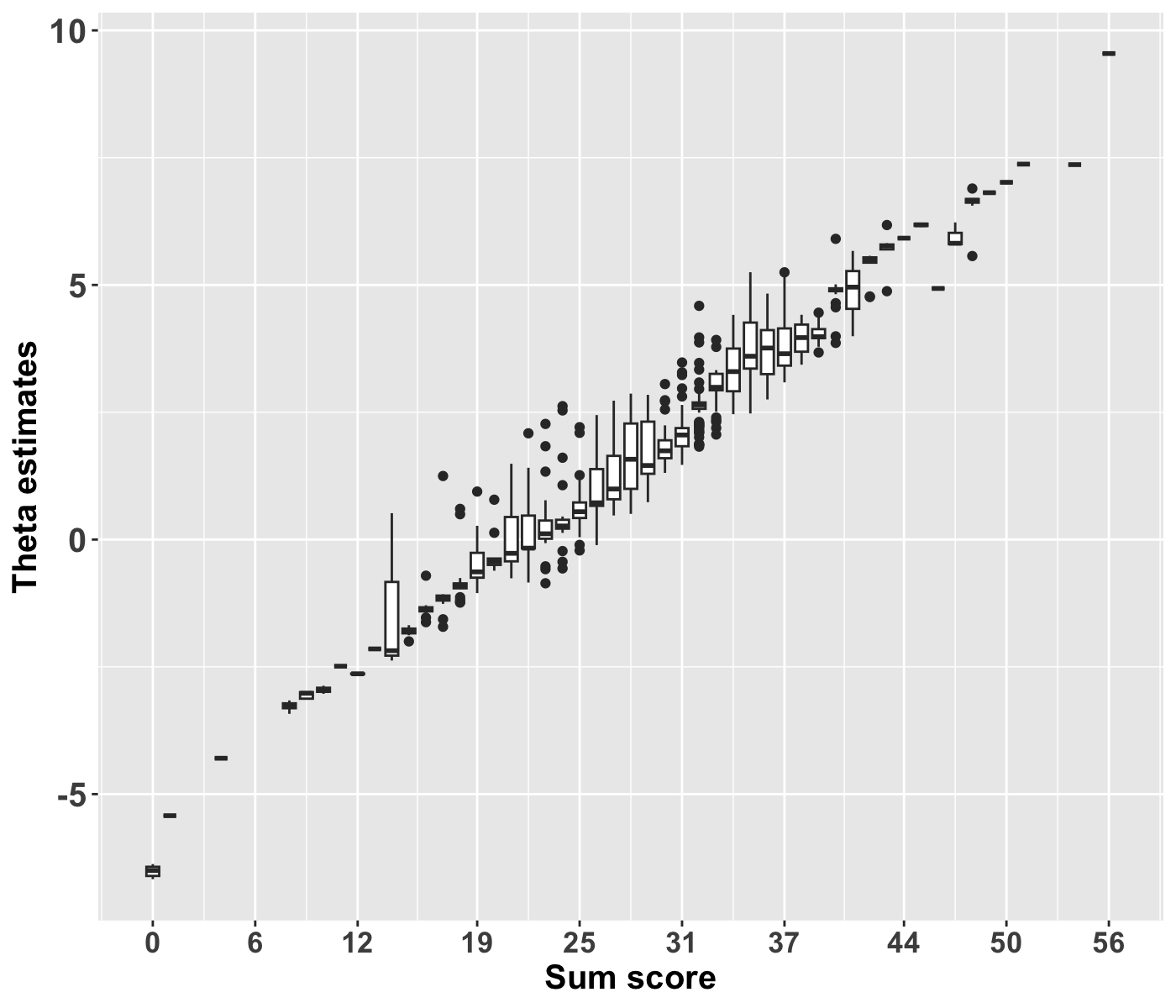}
        \caption{$\theta_k$} \label{fig:latent_paramb}
    \end{subfigure}
    \caption{
    Visual summaries of the parameter estimate of $\beta_i$ and $\theta_k$ using the \code{plot()} function based on the 1PL LSIRM fitted to the TDRI dataset. (a) Boxplots of the posterior samples for $\beta_i$; outliers are suppressed in the boxplots for the sake of simplicity. (b) Boxplots of the point estimate for $\theta_k$ as a function of the total sum scores of positive responses.
    }
    \label{fig:latent_param}
\end{figure}

Figures \ref{fig:latent_param} display the \code{plot()} results for summarizing $\beta_i$ and $\theta_k$ by setting the \code{option} argument as the \code{"beta"} and \code"{theta"}, respectively. \code{option="beta"} generates the boxplots of the posterior samples for $\beta_i$, while \code{option="theta"} generates the boxplots of the point estimates for $\theta_k$, plotted against the total sum scores of positive responses (ranging from 0 to $P$, where $P$ is the number of items). In Figure \ref{fig:latent_parama}, as the item number increases (from left to right on the x axis), the $\beta_k$ estimates decrease, indicating that earlier items are easier. In Figure \ref{fig:latent_paramb}, individuals who correctly answer more items (i.e., higher sum scores) have higher $\theta_i$ estimates. It is sensible that those who answer more items correctly have higher $\theta_i$ values, as $\theta_i$ represents their ability levels.
% In Figure \ref{fig:latent_parama}, outliers are suppressed in the boxplot for the sake of simplicity. 

%which visualizes unobserved interactions between items and respondents. 
The \code{plot()} also returns the interaction map with \code{option="interaction"}. The interaction map is created based on the estimated latent position of the item and respondent, $\boldsymbol{w}_i$ and $\boldsymbol{z}_k$, respectively. Note that the primary and unique advantage of the LSIRM lies in deriving intuitive information from the interaction map, based on the distances between respondents and items, between respondents, and between items. In the interaction map, a shorter distance between the latent position of the item $i$ ($\boldsymbol{w_i}$) and the latent position of the respondent $k$ ($\boldsymbol{z_k}$) indicates a stronger dependence (or interactions), which implies that the respondent $k$ is more likely to respond correctly (or positively) to the item $i$, given the person's ability. In the \CRANpkg{lsirm12pl} package, the interaction map is set to a two-dimensional space, as in \cite{jeon2021mapping}, for parsimony and interpretability. 
\begin{example}
R > plot(lsirm_result, option = "interaction", chain.idx = 1)
\end{example}
\begin{figure}[ht]
     \centering
     \begin{subfigure}[b]{0.4\textwidth}
        \centering
        \includegraphics[width=\textwidth]{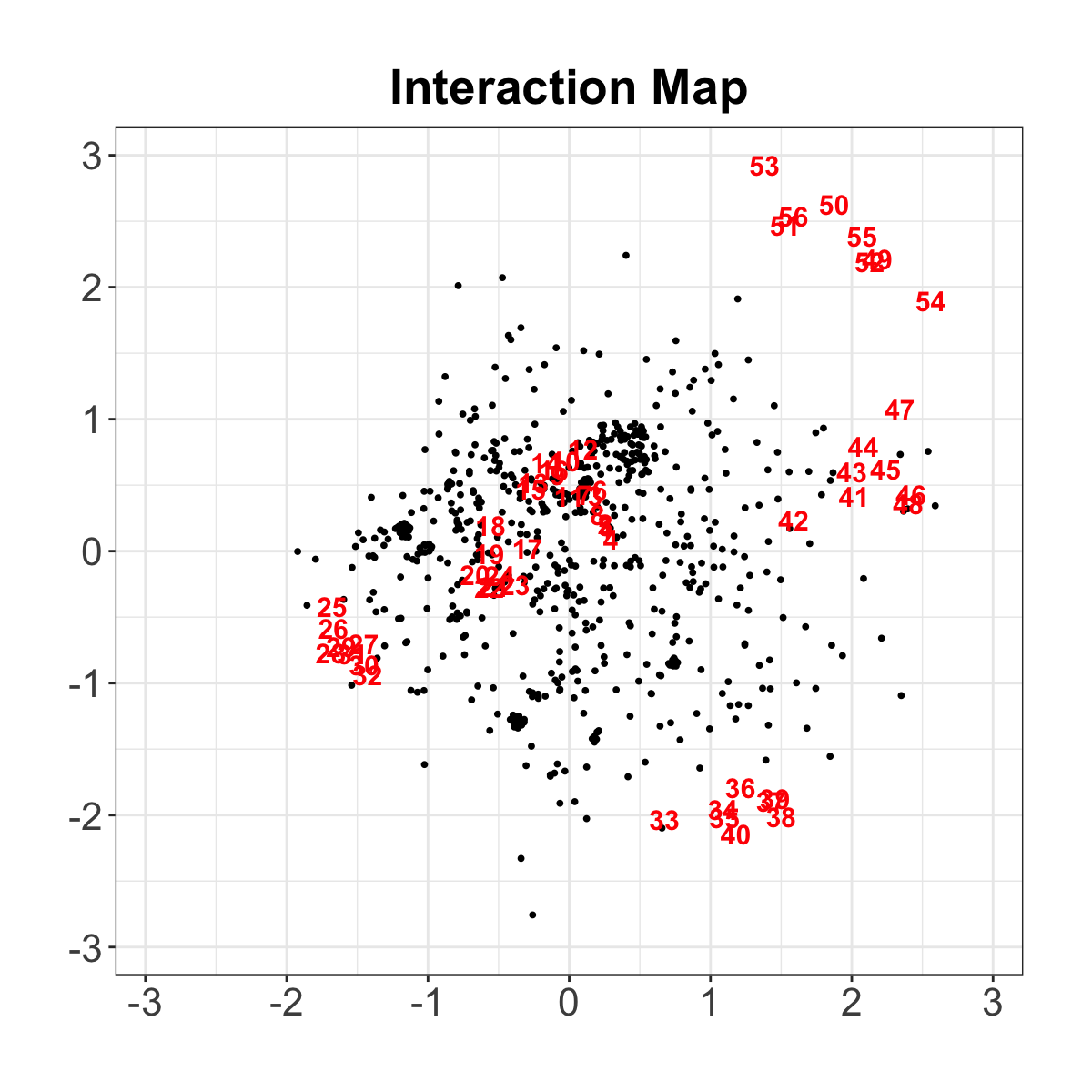}
        \caption{Original} \label{fig:1pl_latent_mapa}
     \end{subfigure}
     \hspace{0.03\textwidth}
     \begin{subfigure}[b]{0.4\textwidth}
        \centering
        \includegraphics[width=\textwidth]{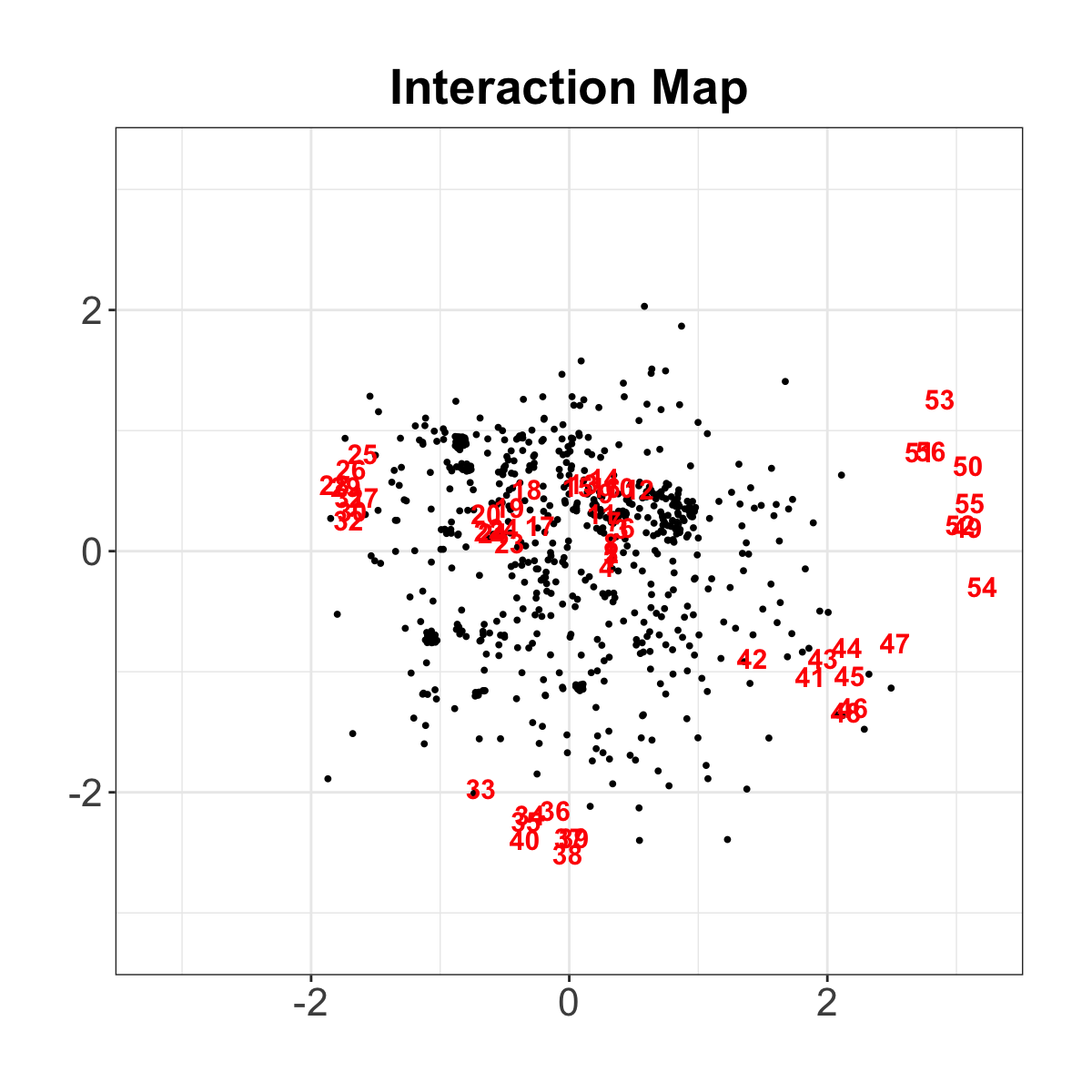}
        \caption{Oblimin rotation} \label{fig:1pl_latent_mapb}
     \end{subfigure}\\
     \begin{subfigure}[b]{0.4\textwidth}
        \centering
        \includegraphics[width=\textwidth]{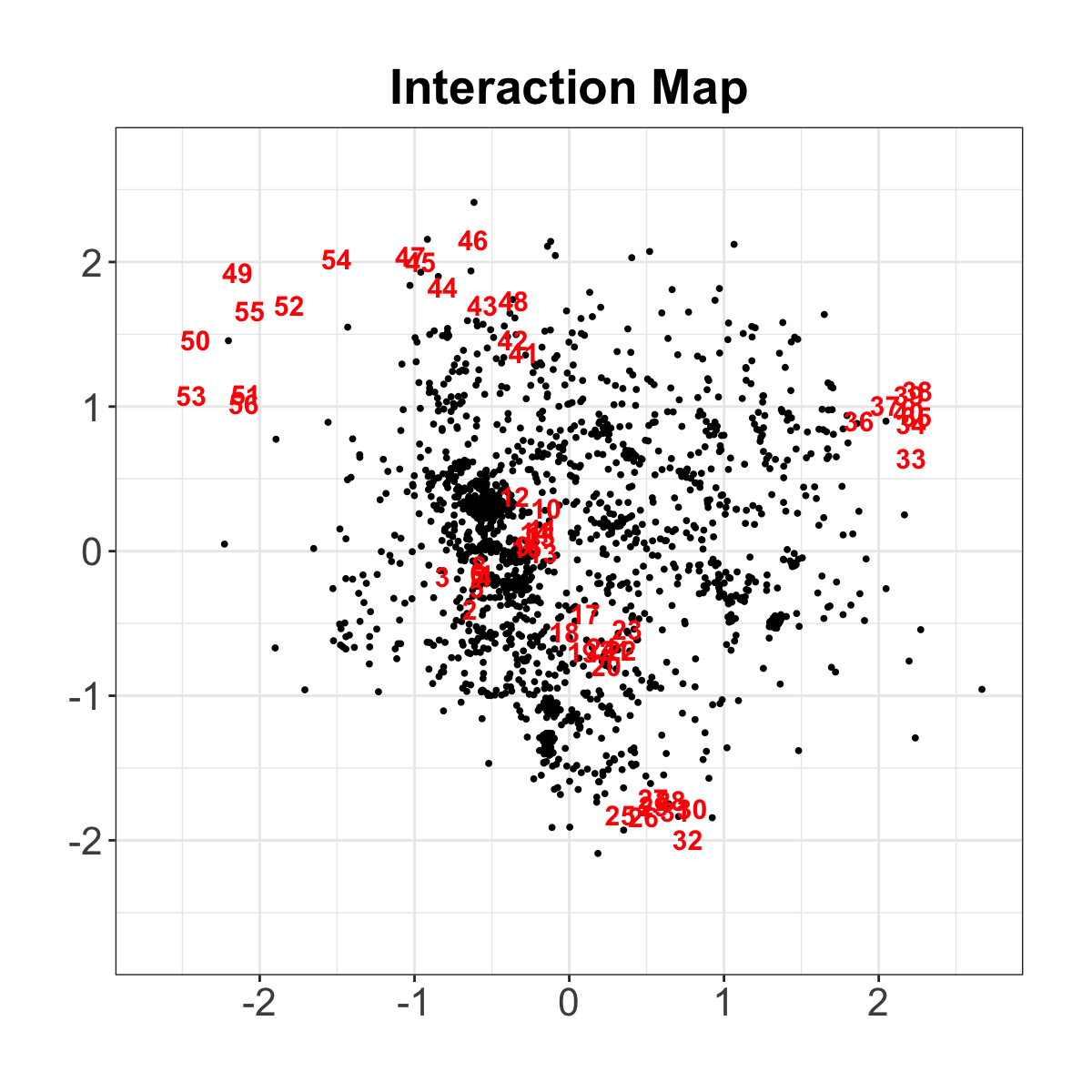}
        \caption{MAR} \label{fig:1pl_latent_mapc}
     \end{subfigure}
     \hspace{0.03\textwidth}
     \begin{subfigure}[b]{0.4\textwidth}
        \centering
        \includegraphics[width=\textwidth]{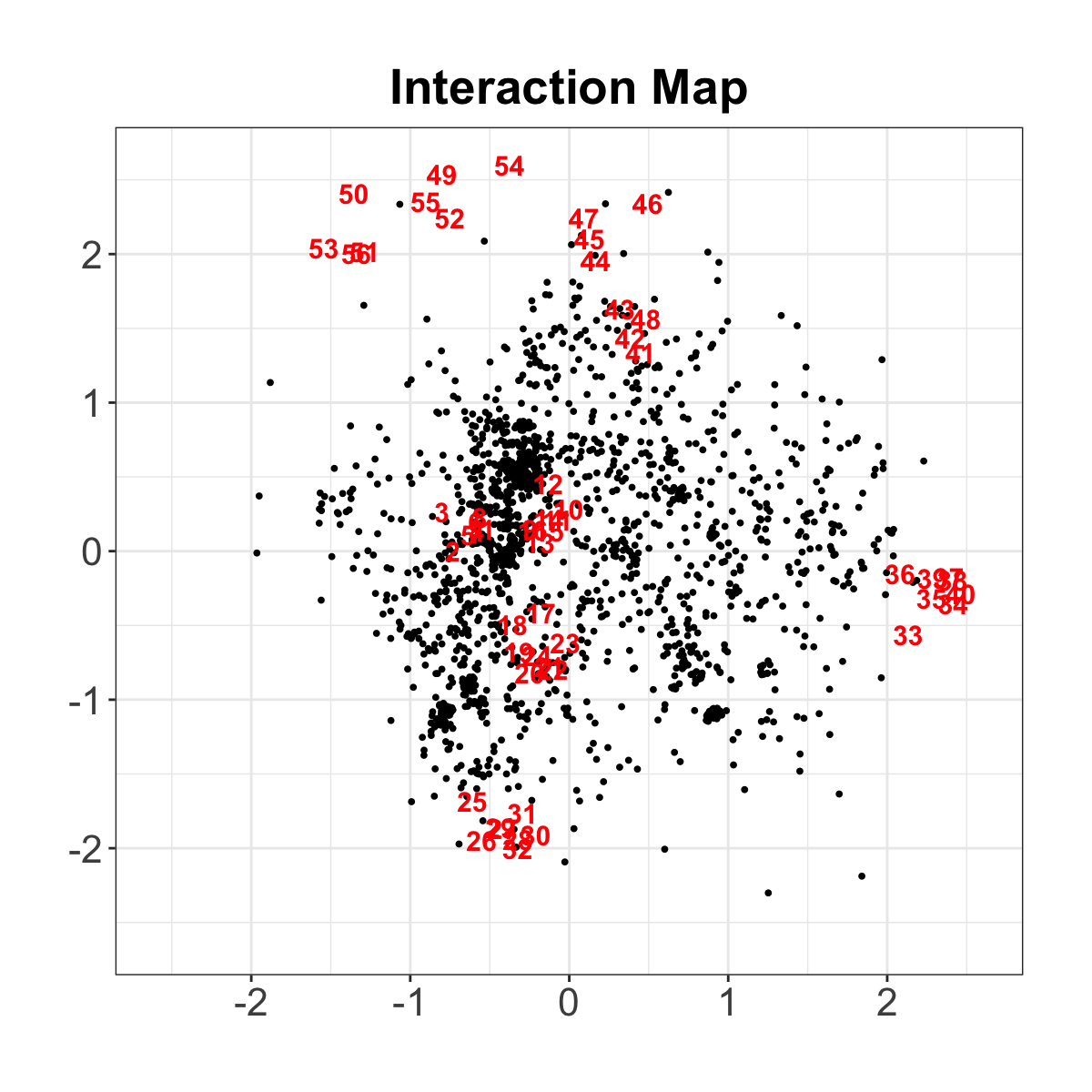}
        \caption{MCAR} \label{fig:1pl_latent_mapd}
    \end{subfigure}
    \caption{
    The estimated interaction maps based on the 1PL LSIRM fitted to the TDRI dataset. (a) The original interaction map. (b) The rotated interaction map using oblimin rotation. (c) The interaction map based on the MAR assumption. (d) The interaction map based on the MCAR assumption which has been reflected across the y-axis to facilitate comparisons with the other interaction maps. Red numbers represent item positions and black dots represent respondent positions in all plots. 
    }
    \label{fig:1pl_latent_map}
\end{figure}

Figure \ref{fig:1pl_latent_mapa} (top left) displays the interaction map based on the 1PL LSIRM for the TDRI data. The latent positions of items are represented as red numbers and respondents as black dots on the map. Note that the two coordinates (dimensions) of the map do not represent specific quantities or substantive meaning, as the interaction map is simply a tool to represent respondent-by-item interactions \citep{jeon2021mapping}. In Figure \ref{fig:1pl_latent_mapa}, we observe that respondents are widely spread around the center of the interaction map, while the items are  separated by some clusters. For example, items located in the far north, such as items 50, 52, 53, 54, and 55 (red numbers) are far from most respondents (black dots) on the map. This suggests that most respondents are less likely to respond correctly to those specific items, regardless of their ability levels. In other words, those items located in the south are difficult items, which is also confirmed in Figure \ref{fig:latent_param}(a). 

\begin{example}
R > plot(lsirm_result, option = "interaction", rotation = TRUE, chain.idx = 1)
\end{example}

If desired, one can rotate the interaction map to improve the interpretability of the coordinates of the map 
with \code{rotation=TRUE}. 
%by setting \code{rotation} option as \code{TRUE}.
The \CRANpkg{lsirm12pl} package offers the oblimin rotation \citep{Jennrich2002} using the \CRANpkg{GPArotation} package in R \citep{Bernaards2005}. Figure \ref{fig:1pl_latent_mapb} (top left) shows the rotated version of the original map shown in Figure \ref{fig:1pl_latent_mapa} (top right). In this example, the original and rotated maps appear pretty similar, although on the right there seems to be slight clockwise rotation compared to the one on the left; after rotation, the x- and y- coordinates may be interpreted based on the items that are closely placed to the respective coordinates. %  this may be because the items are already being placed close to the two coordinates in Figure \ref{fig:1pl_latent_mapa}.  

Further, the \CRANpkg{lsirm12pl} package offers an option to cluster latent positions of items. Two types of clustering methods are available: spectral clustering \citep{Ng2001, vonLuxburg2007} and the Neyman-Scott process modeling approach \citep{Thomas1949, Neyman1952, yi2023}, with \code{cluster} option as \code{spectral} and \code{neyman}, respectively. Spectral clustering is a technique that clusters points based on the eigenvalues of a similarity matrix, using the spectral properties of the data to identify clusters. The implementation of spectral clustering is based on the \CRANpkg{specc} package \citep{Karatzoglou2004} in R where the number of clusters is determined using the average silhouette width \citep{Batool2021}. The Neyman-Scott point process modeling approach is a method to cluster points in time or space. Parent points (cluster center) are generated using a Poisson process, with offspring points clustered around each parent based on a defined probability distribution. The \CRANpkg{lsirm12pl} package implements this method directly; it applies the MCMC algorithm a number of times independently to determine the distribution of the cluster number and cluster centers. Then, the mode of the cluster number distribution is chosen as the optimal cluster number. With the optimal cluster number, the cluster center that minimizes the Bayesian information criterion is selected, and items are assigned to the nearest cluster based on Euclidean distances. 

% \textcolor{red}{[[mj: it would be helpful to provide a brief explanation of both clustering methods. Also, what do you mean by "apply the MCMC algorithm 100 times?"]]}

\begin{example}
R > plot(lsirm_result, cluster = "spectral", chain.idx = 1)

Clustering result (Spectral Clustering): 
  group                                      item 
      A              49, 50, 51, 52, 53, 54, 55, 56        
      B              33, 34, 35, 36, 37, 38, 39, 40 
      C              41, 42, 43, 44, 45, 46, 47, 48       
      D              1, 2, 3, 4, 5, 6, 7, 8, 9, 10,
                     11, 12, 13, 14, 15, 16, 17, 18,
                     19, 20, 21, 22, 23, 24
      E              25, 26, 27, 28, 29, 30, 31, 32

R > plot(lsirm_result, cluster = "neyman", chain.idx = 1)

|==================================================| 100%
Clustering result (Neyman-Scott process): 
  group                                      item 
      A             33, 34, 35, 36, 37, 38, 39, 40 
      B             49, 52, 54, 55  
      C             25, 26, 27, 28, 29, 30, 31, 32 
      D             17, 18, 19, 20, 21, 22, 23, 24 
      E             50, 51, 53, 56 
      F             9, 10, 11, 12, 13, 14, 15, 16 
      G             1, 2, 3, 4, 5, 6, 7, 8 
      H             41, 42, 43, 44, 45, 46, 47, 48 
\end{example}

%The results  spectral clustering and the Neyman-Scott process model are shown below on the R console. 

\begin{figure}[ht]
    \centering
    \begin{subfigure}[b]{0.35\textwidth}
        \centering
        \includegraphics[width=\textwidth]{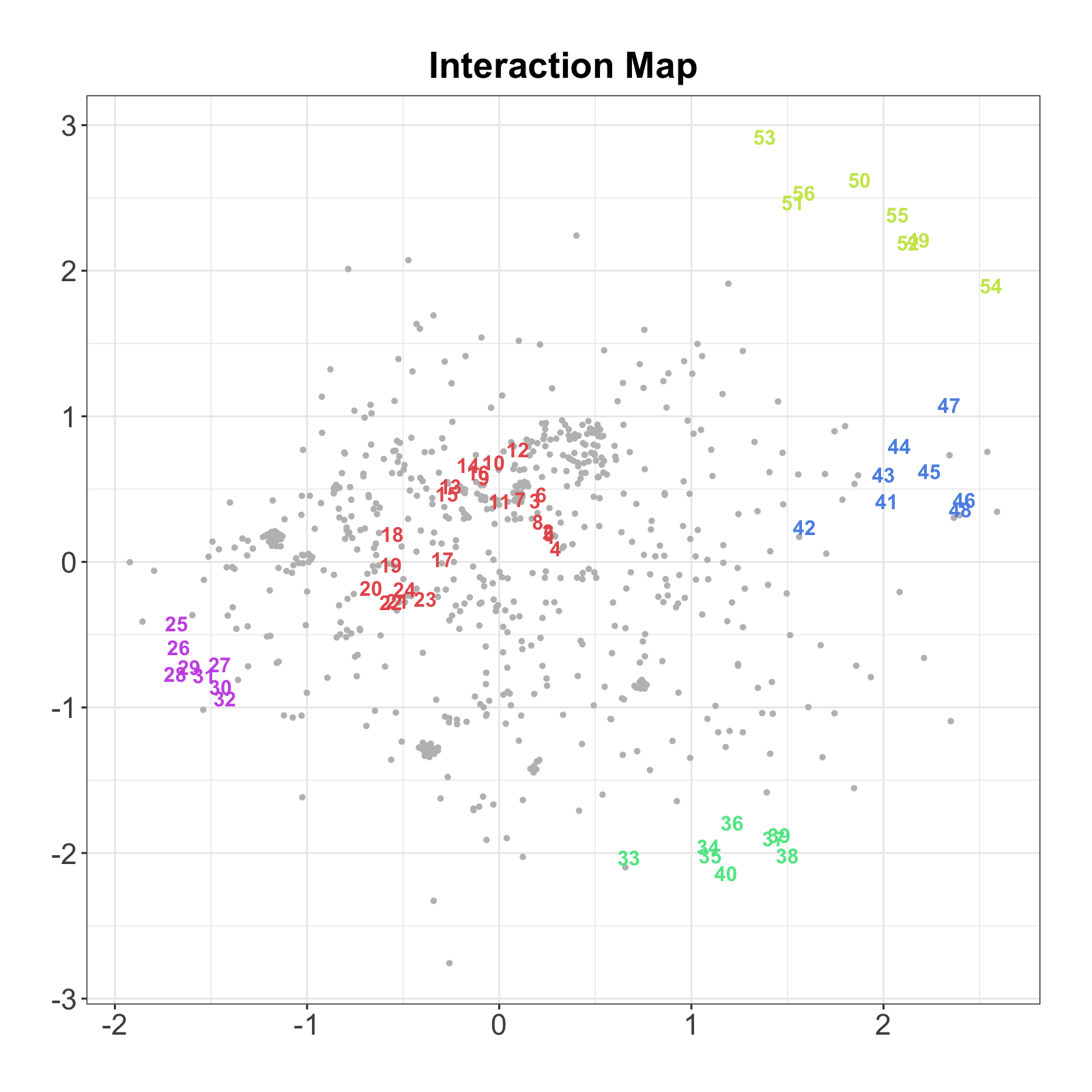}
        \caption{Spectral Clustering} \label{fig:1pl_latent_map_clustera}
    \end{subfigure}
    \hspace{0.05\textwidth}
    \begin{subfigure}[b]{0.35\textwidth}
        \centering
        \includegraphics[width=\textwidth]{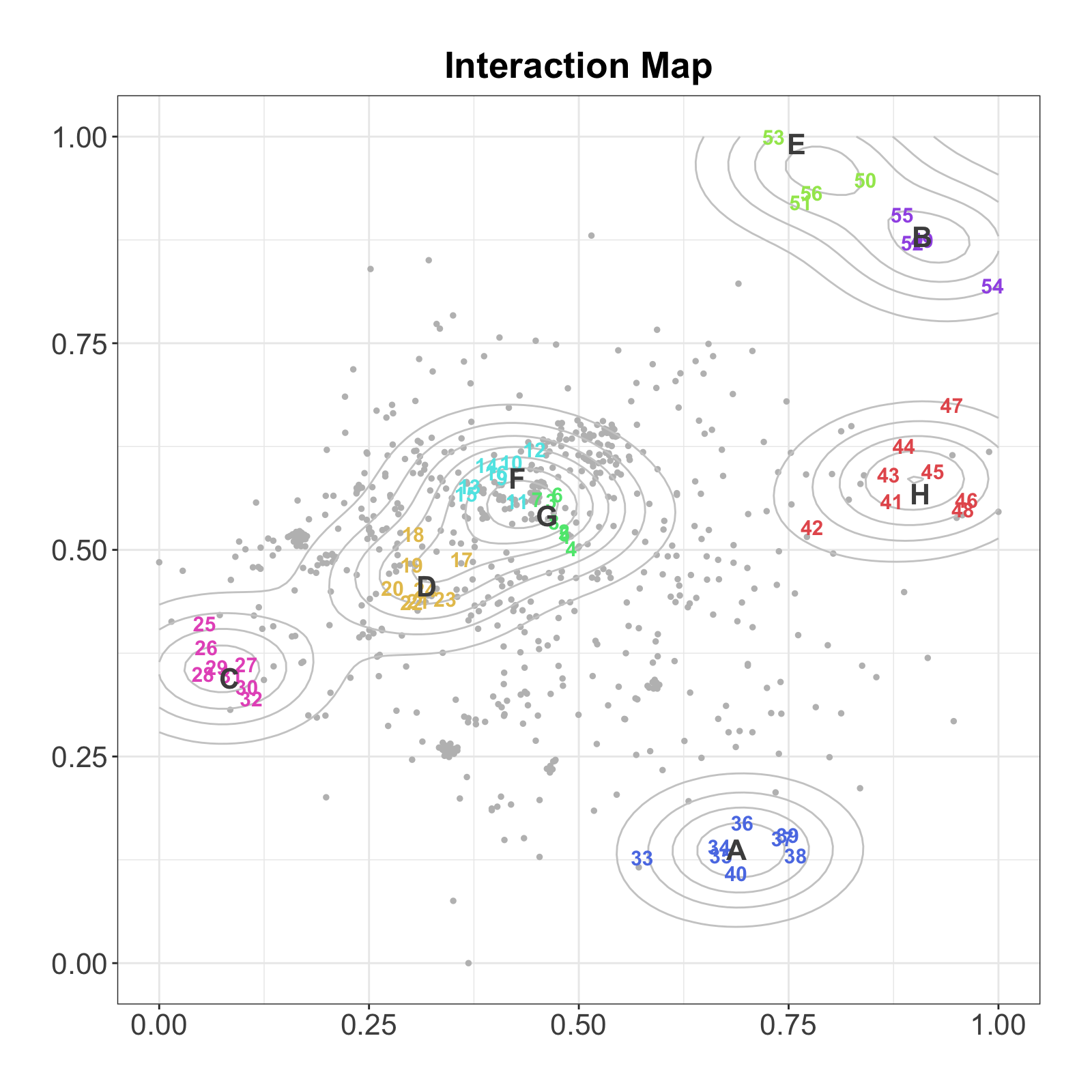}
        \caption{Neyman-Scott Process Model} \label{fig:1pl_latent_map_clusterb}
    \end{subfigure}
    \caption{
    The interaction map with item clustering results based on the 1PL LSIRM fitted to the TDRI dataset, using (a) spectral clustering and the (b) Neyman-Scott process modeling approach. In both plots, the gray dots indicates respondents, while   numbers in  colors indicate items with different cluster memberships. The Neyman-Scott process approach additionally displays the center of the cluster (alphabets A to H) and a contour for each cluster. 
    }
    \label{fig:1pl_latent_map_cluster}
\end{figure}

Figures \ref{fig:1pl_latent_map_clustera} and \ref{fig:1pl_latent_map_clusterb} are item clustering results based on spectral clustering and the Neyman-Scott process model, respectively. In both plots, the gray dots indicate respondents, where numbers in colors indicate items with different cluster memberships. The Neyman-Scott process modeling approach additionally displays the center of the item cluster (alphabets A to H) and a contour for each item cluster. In this example, the spectral clustering method identifies five clusters, while the Neyman-Scott process modeling approach identifies eight clusters. Overall, the Neyman-Scott process modeling approach further split the items near the center (red items on the left) and in the north (light green items on the left) compared to the spectral clustering method. 
%In both results, the items with similar indices are grouped together in clusters. The Neyman-Scott process modeling approach shows more detailed item clustering compared to spectral clustering. 

% \textcolor{red}{[[mj: please add a brief explanation of these results. why the results are different and which results should one follow? ]]}

\subsection{Flexible Modeling Options}\label{1pl_extension}

The \CRANpkg{lsirm12pl} package provides a range of flexible modeling options for the LSIRM. 

First, with \code{fixed\_gamma} option, one can fix the distance weight $\gamma$ to a constant value of 1. By doing so, the scale of the latent space can be standardized, easing to compare different interaction maps. The default value of \code{fixed\_gamma} option is \code{FALSE} (i.e.,  $\gamma$ is freely estimated).
%To implement this option, the main function offers a Boolean input called \code{fixed\_gamma}, which can be set to TRUE if the distance weight $\gamma$ is to be fixed. 
%The following is the example code for fitting the 1PL LSIRM with a fixed $\gamma$ by specifying \code{fixed\_gamma} as \code{TRUE}. \textcolor{blue}

\begin{example}
R > lsirm_result <- lsirm(data ~ lsirm1pl(fixed_gamma = TRUE))
\end{example}

Second, with \code{spikenslab} option, one can apply a model selection with the spike-and-slab prior for the distance weight $\gamma$ \citep{Ishwaran2005}. The spike-and-slab prior is a mixture of two log-normal priors: one is densely concentrated near zero, while another is more broadly spread across positive values. With this option, one can determine whether $\gamma$ is zero or not, which in turn determines whether the distance term is needed for the data being analyzed. If $\gamma$ is not zero, it is evidence for item-by-respondent interactions in the data being analyzed, and thus it  would be useful to investigate the interactions between respondents and items with the LSIRM approach. The default value of \code{spikenslab} option is \code{FALSE}.
%\textcolor{red}{[[mj: a brief explanation of the spike and slab prior]]}

The posterior probability of $\gamma$ being non-zero is indicated by the return value of \code{pi\_estimate}. A \code{pi\_estimate} value greater than 0.5 suggests that $\gamma$ is likely non zero. Note that the two options \code{fixed\_gamma} and \code{spikenslab} cannot be used simultaneously. %\textcolor{blue}{For simplicity, we fit a single MCMC chain by setting the \code{chain} option to 1.}
%The spike-and-slab prior is included in the package as an option with \code{spikenslab}. 

\begin{example}
R > lsirm_result <- lsirm(data ~ lsirm1pl(spikenslab = TRUE))
R > lsirm_result$pi_estimate
[1] 0.9984
\end{example}

Third, the package offers options for handling missing data. In the context of missing data, three key assumptions are considered \citep{rubin1976inference}: missing completely at random (MCAR), missing at random (MAR), and missing not at random (MNAR). MCAR occurs when the probability of missing data on a variable is unrelated to any other measured variables or the variable itself, making the missingness entirely random. MAR happens when the probability of missing data on a variable is related to some of the observed data but not the missing data itself, meaning the missingness can be explained by other observed variables. Unlike MCAR and MAR, MNAR assumes that the missing data mechanism depends on the unobserved data itself, making it challenging to estimate the model without strong assumptions or additional information about the missing data. Therefore, we focus on two types of missing data, MCAR and MAR, in the \CRANpkg{lsrim12pl} package with the \code{missing\_data} option.
%On the other hand, MNAR occurs when the probability of missing data on a variable is related to the unobserved data itself, meaning the missingness is related to the missing values. 

With \code{missing\_data = "mcar"}, the missing data are assumed to follow MCAR and the parameters are estimated solely based on the observed elements of the dataset being analyzed. On the other hand, with \code{missing\_data="mar"}, the missing data are assumed to be MAR, and the data augmentation algorithm \citep{Tanner:1987p528} is applied to impute missing values. With \code{missing\_data="mar"}, the function returns the posterior samples of the imputed responses with \code{imp} and the probability of a correct response with \code{imp\_estimate}. Imputed values are listed in the order of respondents. Note that all missing values should be recoded via \code{missing.val}. The percentage of missing data in the TDRI dataset is 30\%, and missing values are replaced with 99 which is the default coding of missing data. The \code{missing\_data} option can be used in combination with other options such as \code{(spikenslab = TRUE, missing\_data = "mar").} 
% with the default value of 99. 
%\textcolor{blue}{For simplicity, we fit the single MCMC chain by setting \code{chain} option as 1.}

\begin{example}
R > data <- lsirm12pl::TDRI
R > data[is.na(data)] <- 99
R > lsirm_result <- lsirm(data ~ lsirm1pl(missing_data = "mcar"))
R > lsirm_result <- lsirm(data ~ lsirm1pl(missing_data = "mar"))
R > lsirm_result$imp_estimate
[1] 0.9900 0.9868 0.9768 0.9884 0.9716 0.9716
...
[997] 0.9860 0.9788 0.8240 0.9924
 [ reached getOption("max.print") -- omitted 32327 entries ]
R > plot(lsirm_result, option = "interaction")
\end{example}

Figures \ref{fig:1pl_latent_map}(c) and \ref{fig:1pl_latent_map}(d) show the resulting interaction maps with MAR and MCAR assumptions, respectively. Note that a reflection is applied to the interaction map with MCAR assumptions (c) across the y-axis to ease comparisons with the other interaction plots. Reflection or rotation does not change the interpretations of the interaction map as interpretations are based on the distances, not the positions themselves. If the interaction maps of two missing assumptions are considerably different, further investigation is necessary to determine which assumption would be more suitable for the analyzed data. In this example, the interaction maps with the missing data options are similar to the original map presented in Figure \ref{fig:1pl_latent_mapa}. Note that the original interaction map (a) is based on the complete item response data with no missing values; that is, the data includes only respondents who answered all items. In contrast, the interaction maps with the missing data option are based on MAR and MCAR assumptions. The similarity between these maps suggests that the observed only or imputed item response data provides a reasonable representation of the original item response data. 

% \textcolor{red}{[[(1)mj: it would be better to rotate the plots, so that they can look similar to each other and to the original plots. Right now, they do not look similar. (2) Also, what does it mean that the two missing data assumptions are similar? (3) it might be useful to report the missing data percentage - 30\%]]}

% \textcolor{red}{[[mj: what do you mean by "complete item response data"? how is it different from mcar assumption?]]}
% %, whereas the maps based on various missing assumptions were created from the original data, including respondents with missing item responses. 

%\textcolor{red}{[[is this a reasonably interpretation? related to my comments above, I would add a bit more about]] }

\section{2PL LSIRM For Dichotomous Data}\label{2pl}

The two-parameter LSIRM (2PL LSIRM) extends the 1PL LSIRM with item discrimination parameters. %as shown in Equation \eqref{birnbaum2pl_formula}. 

\subsection{Statistical framework}

The 2PL LSIRMs the probability of a correct response by respondent $k$ to item $i$ as follows:
\[
\text{logit}\Big(\mathbb{P}(Y_{k, i}=1 | \theta_{k}, \alpha_{i}, \beta_{i}, \gamma, \boldsymbol{z_{k}, w_{i}})\Big) = \alpha_{i} \theta_{k}+ \beta_{i} - \gamma d (\boldsymbol{z_{k}}, \boldsymbol{w_{i}}), \; \; \theta_{k} \sim N(0, \sigma^2), 
\]
where $\theta_k$, $\beta_i$, ${\bf z}_k$, ${\bf w}_i $ and $\gamma$ have similar interpretations to Equation \eqref{1pl_formula}, while $\alpha_i$ represents the item discrimination parameters and one of the $\alpha_i$ parameters is fixed at 1, e.g., $\alpha_{1} =1$ to ensure identifiability. 
% \textcolor{red}{mj: please confirm if that is what is done in the package/ JA: done}

The observed data likelihood function under the 2PL LSIRM is given as
\begin{align*}
\mathbb{L}\left(\boldsymbol{Y=y|\theta,\alpha,\beta,\gamma,Z,W}\right)=\prod_{k=1}^{N}\prod_{i=1}^{P}\mathbb{P}(Y_{k,i}=y_{k,i}|\theta_{k},\alpha_{i}, \beta_{i},\gamma,\boldsymbol{z_{k},w_{i}}).
\label{likelihood_2pl}
\end{align*}
\vspace*{-1cm}

\subsection{Parameter Estimation}\label{2pl-esti}

Following is the posterior distribution of the 2PL LSIRM.
\begin{align*}
\pi\left(\boldsymbol{\theta, \beta}, \gamma, \boldsymbol{Z, W} | \boldsymbol{Y=y} \right) & \propto \prod_{k=1}^{N}\prod_{i=1}^{P}\mathbb{P}(Y_{k,i}=y_{k,i}|\theta_{k},\alpha_{i}, \beta_{i},\gamma,\boldsymbol{z_{k},w_{i}}) \\
&\times \prod_{k=1}^N \pi(\theta_k) \times \prod_{i=1}^P \pi(\alpha_i) \times \prod_{i=1}^P \pi(\beta_i) \times \pi(\gamma) \times \prod_{k=1}^N \pi(\boldsymbol{z_{k}}) \prod_{i=1}^P \pi(\boldsymbol{w_{i}})
\end{align*}
The prior distributions for $\theta_k$, $\beta_i$, ${\bf z}_k$, ${\bf w}_i$, and $\gamma$ are identical for the 1PL LSIRM. For item discrimination parameters $\alpha$, a log-normal distribution is used with a mean of $\mu_{\alpha}$ and a variance of $\tau_{\alpha}^2$, as $\alpha$ is typically assumed to be positive. The argument names and default values for the prior specification are shown in the Github site. For the item discrimination parameters $\alpha$, the arguments and default values are \code{pr\_mean\_alpha = 0.5, pr\_sd\_alpha = 1}. 
%To estimate the parameters in the 2PL LSIRM, a fully Bayesian approach with Markov chain Monte Carlo (MCMC) as we did with the 1PL LSIRM. 
% since item discrimination parameters $\alpha$ are typically assumed to be positive, 
% the same for the parameters $\theta$, $\beta$, $\gamma$, $\sigma$ as described in Table \ref{prior_option} in the appendix. 

The conditional posterior distribution for each parameter follows the same form as the Equation in the Github site. The jumping rule for each parameter is given in the Github site. The default jumping rule for $\alpha$ is \code{jump\_alpha = 1}. 

\vspace{.1 in}

\subsection{An Illustrated Example} \label{2pl_extension}

We apply the 2PL LSIRM to the TDRI data. The default settings of the 2PL LSIRM are the same as the 1PL LSIRM except for $\alpha$. The base function for the 2PL LSIRM is 
\begin{center}
\code{lsirm(A $\sim$ lsirm2pl())}    
\end{center}

The 2PL LSRIRM for dichotomous data was fitted with 4 MCMC chains using 2 multi-core processors, as demonstrated in the following example code. Similarly to the \code{lsirm1pl} results, the estimation results for the model parameters $\boldsymbol{\theta,\ \beta,\ \alpha}$, $\gamma,\ \boldsymbol{Z}$, and $\boldsymbol{W}$ for each chain are provided in individual lists.
%This function returns a list of estimates on the model parameters, $\boldsymbol{\theta,\ \beta,\ \alpha}$, $\gamma,\ \boldsymbol{Z}$, and $\boldsymbol{W}$. 

\begin{example}
R > library("lsirm12pl")
R > data <- lsirm12pl::TDRI
R > data <- data[complete.cases(data),]
R > lsirm_result <- lsirm(data ~ lsirm2pl(chains = 4, multicore = 2, seed = 2025))
\end{example}

% \begin{example}
% R > summary(lsirm_result)
% ========================== 
% Summary of model 
% ========================== 

% Call:	 lsirm.formula(formula = data ~ lsirm2pl()) 
% Model:	 2pl LSIRM 
% Data type:	 binary 
% Variable Selection:	 FALSE 
% Missing:	 NA 
% MCMC sample of size 15000, after burnin of 2500 iteration 

% Covariate coefficients posterior means:  

%      Estimate      2.5%   97.5%
% i1   8.428021  7.888651  9.0083
% ...
% i56 -0.235174 -0.789505  0.2930

% --------------------------- 

% Overall BIC (Smaller is better) : 49174.75 

% Maximum Log-posterior Iteration:  
%        value iter
% [1,] -9658.7 1622
% \end{example}

To diagnose the results of the 2PL LSIRM analysis, we can use the \code{diagnostic()} function. Similarly to the diagnostic of the 1PL LSIRM, the convergence of MCMC for each parameter can be checked using various diagnostic tools, such as trace plots, posterior density distributions, autocorrelation functions (ACF), and Gelman-Rubin-Brooks plots. By setting the \code{draw.item} option to \code{list('beta'= c(1))}, we perform diagnostics for the $\beta_i$ parameter of the first item ($i=1$).

\begin{example}
R > diagnostic(lsirm_result, 
               draw.item = list(beta = c(1)),
               gelman.diag = T)
               
Potential scale reduction factors:

          Point est. Upper C.I.
beta [i1]          1          1
\end{example}
\begin{figure}[htb]
    \centering
    \includegraphics[width = .8\textwidth]{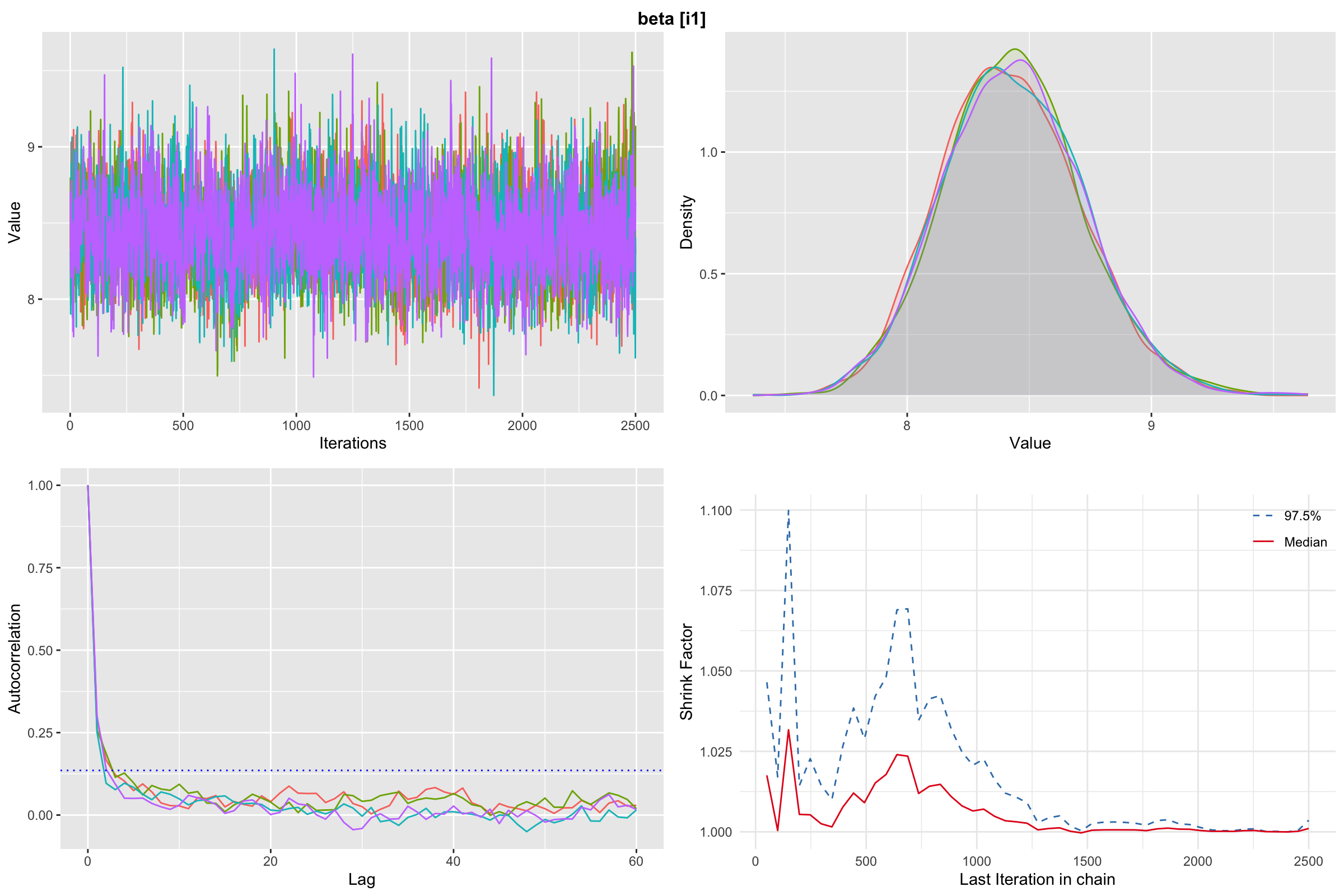}
    \caption{
    Diagnostics of $\beta_1$ using the \code{diagnostic()} function on the results of the 2PL LSIRM fitted to the TDRI dataset. The top left is the trace plot, the top right is the posterior density plot, the bottom left is the autocorrelation plot, and the bottom right is the Gelman-Rubin-Brooks plot. Different colors represent different MCMC chains.
    }
    \label{fig:diagnosis2PL}
\end{figure}

Figure \ref{fig:diagnosis2PL} displays the diagnostic results for $\beta_1$ of the results of the 2PL LSIRM fitted to the TDRI dataset. These plots help us assess the convergence of MCMC for $\beta_1$. The interpretation of each plot from the \code{diagnostic()} function is the same as mentioned for the 1PL LSIRM in the previous Section. In Figure \ref{fig:diagnosis2PL}, the convergence of $\beta_1$ was confirmed by various diagnostic tools.

\begin{figure}[htb]
    \centering
    \includegraphics[width=0.7\textwidth]{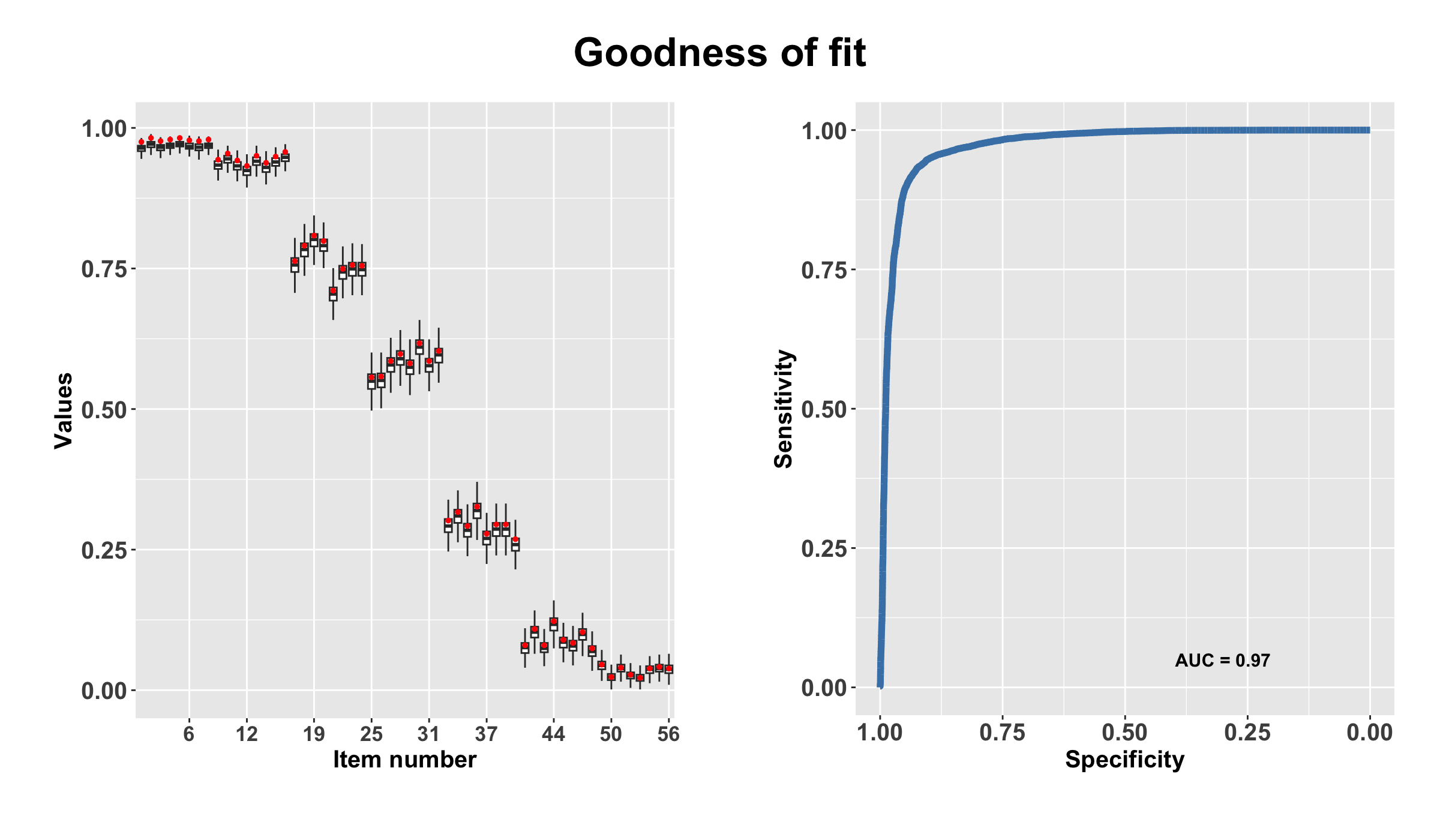}
    \caption{Goodness of fit of the 2PL LSIRM for the TDRI data using the \code{gof()} function. 
    The box plots of the average posterior predicted response values for items against the average observed response value for each item indicated by red dots (left) and the receiver operating characteristic (ROC) curve (right).
    }
    \label{fig:gof2}
\end{figure}

The \code{gof()} function assesses the goodness-of-fit of the LSIRM. Figure \ref{fig:gof2} visualizes the box plots of the posterior predicted response values (item-wise means) compared with the observed item-wise means and the ROC curve to check the performance of the 2PL LSIRM fitted to the TDRI dataset. In the figure, most of the red dots are located close to the midline of the boxplots and the AUC is 0.97, indicating the fit of the 2PL LSIRM to the TDRI dataset is satisfactory. %reasonably well.
%A comprehensive description of this function is provided in the 1PL LSIRM description. 

% \textcolor{red}{[[mj: the fit of items 18-31 do not still look that good. say something about this compared to the 1PL LSIRM case /Solve! (Changes the posterior predicted response value calculation method used when drawing a boxplot.)]]  }

\begin{example}
R > gof(lsirm_result, chain.idx = 1)
\end{example}

Similarly to the 1PL LSIRM, the \code{plot()} function can generate different graphs for the 2PL LSIRM results with the \code{option} argument: 
%, which can be set to \code{"beta"} and \code{"theta"} to summarize the $\beta_i$ and $\theta_k$ parameter estimates, respectively, or to \code{interaction} to display an interaction map.
%The options for this function have been expanded by including the option to draw a boxplot for $\alpha_i$ as well as the options discussed in the 1PL LSIRM description. 

\begin{example}
R > plot(lsirm_result, option = "beta", chain.idx = 1)
R > plot(lsirm_result, option = "theta", chain.idx = 1) 
R > plot(lsirm_result, option = "alpha", chain.idx = 1) 
\end{example}

\begin{figure}[ht]
     \centering
     \begin{subfigure}[b]{0.3\textwidth}
        \centering
        \includegraphics[width=\textwidth]{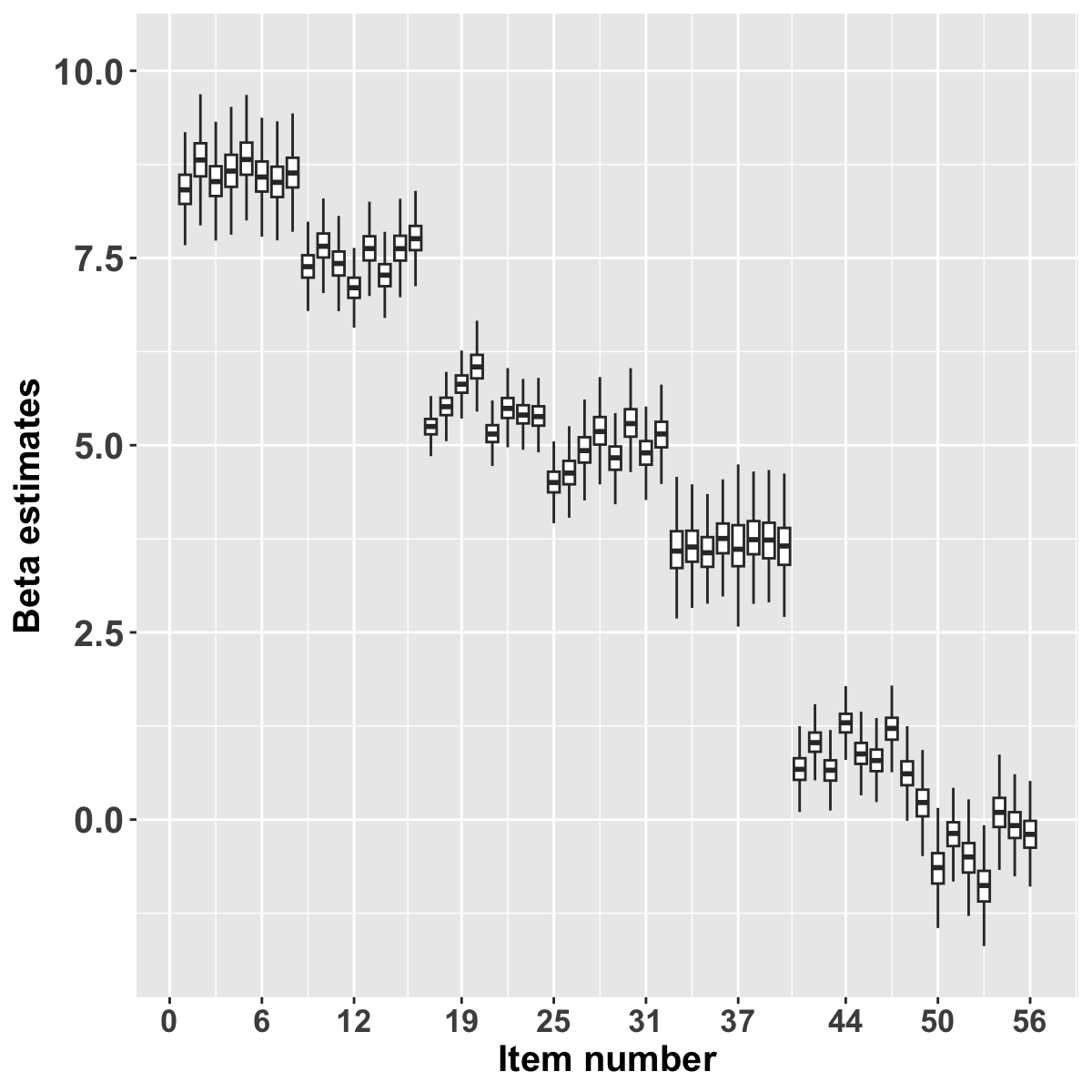}
        \caption{$\beta_i$} \label{fig:latent_param_2pla}
     \end{subfigure}
     \hspace{0.03\textwidth}
     \begin{subfigure}[b]{0.3\textwidth}
        \centering
        \includegraphics[width=\textwidth]{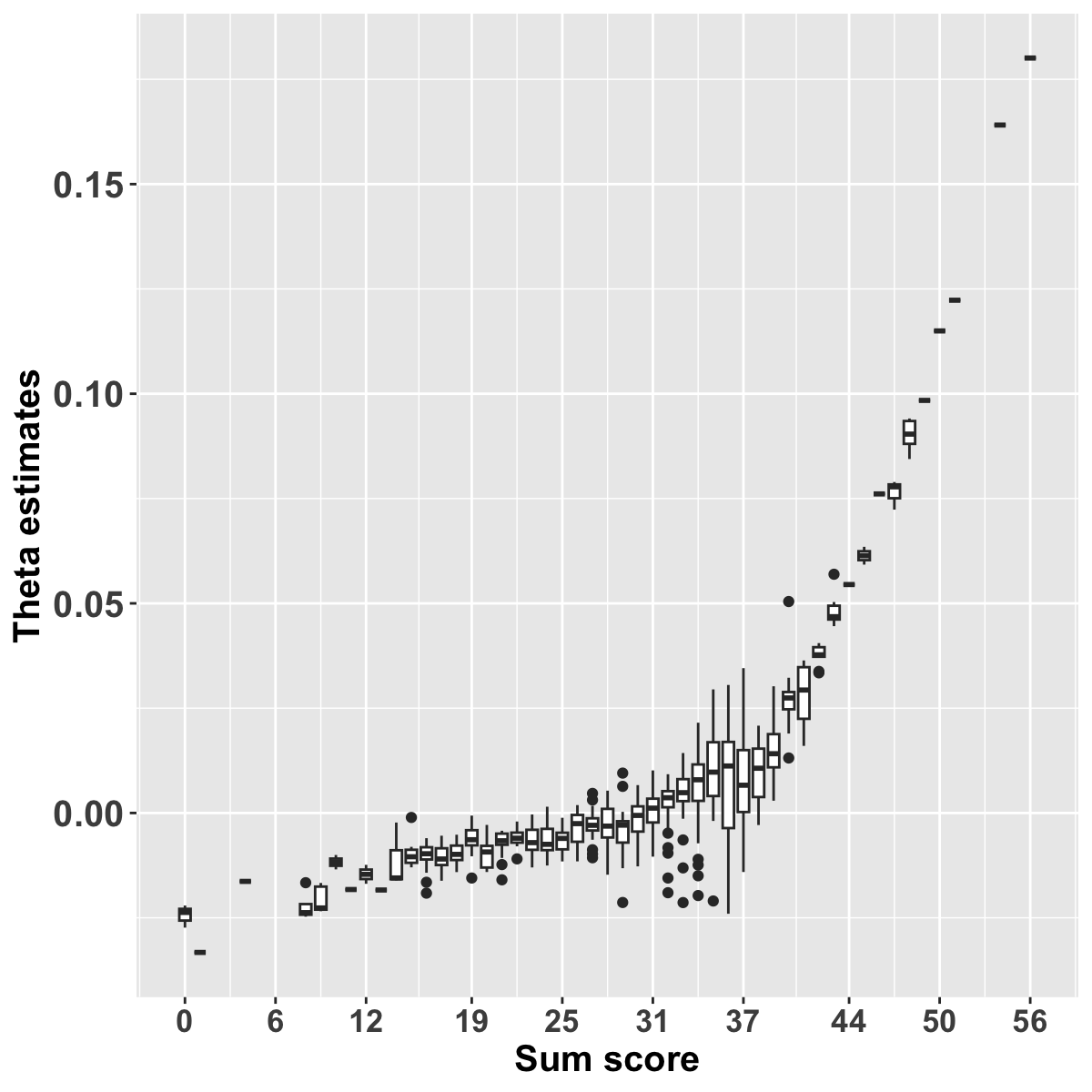}
        \caption{$\theta_k$} \label{fig:latent_param_2plb}
     \end{subfigure}
     \hspace{0.03\textwidth}
     \begin{subfigure}[b]{0.3\textwidth}
        \centering
        \includegraphics[width=\textwidth]{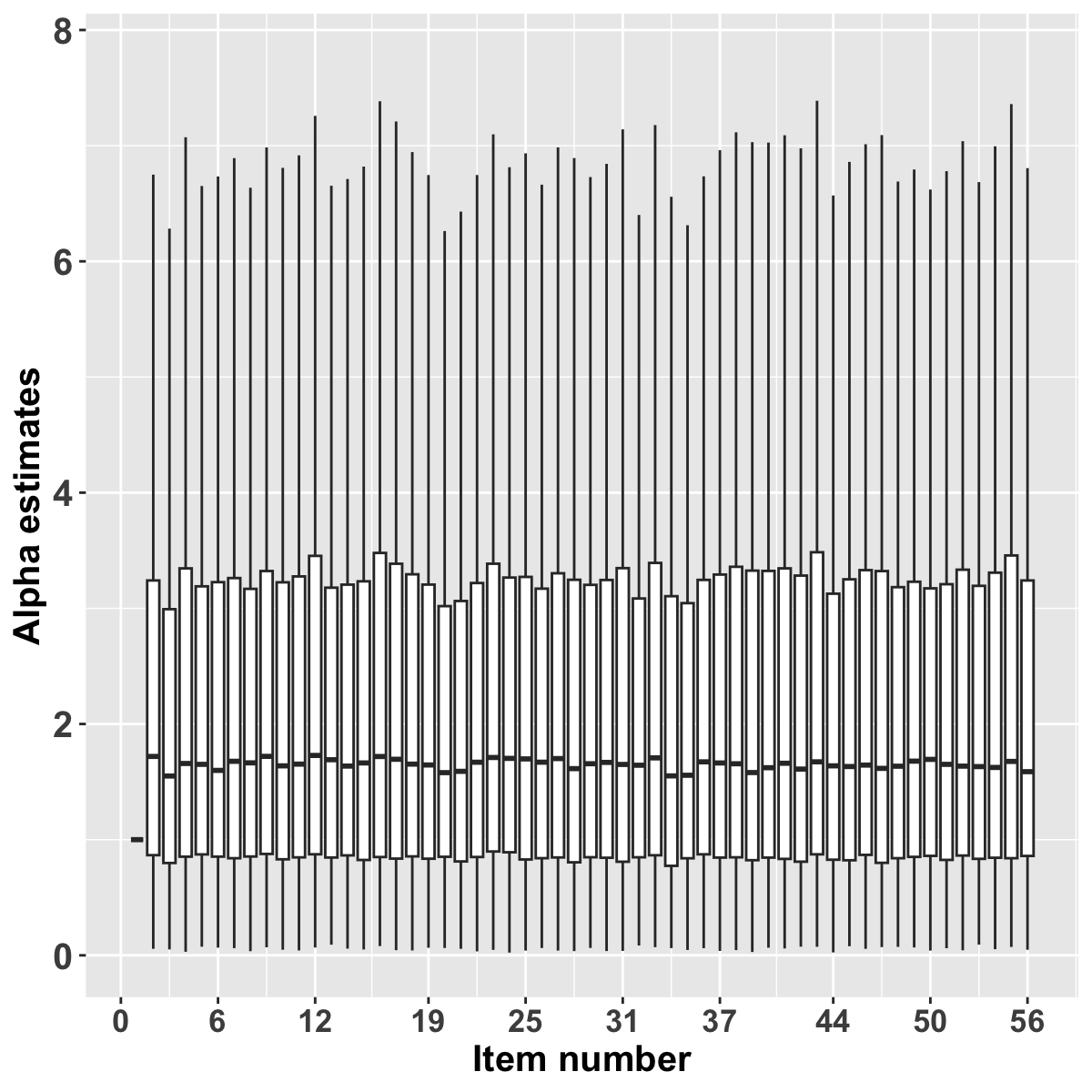}
        \caption{$\alpha_i$} \label{fig:latent_param_2plc}
     \end{subfigure}
      \caption{Summarizing $\beta_i$ and $\theta_k$ using the \code{plot()} function on the results of the 2PL LSIRM fitted to the TDRI dataset. (a) Boxplots of the posterior samples for $\beta_i$. (b) Boxplots of the point estimate for $\theta_k$ as a function of the total sum scores of positive responses. (c) Box plots of the posterior samples for $\alpha_i$.}
    \label{fig:latent_param_2pl}
\end{figure}

Figure \ref{fig:latent_param_2pl} illustrates the results of the \code{plot()} function with the \code{"beta"}, \code{"theta"}, and \code{"alpha"} options for the 2PL LSIRM, respectively. Similarly to the 1PL LSIRM,  $\beta_i$ decreases for later items, indicating that later items are more difficult than earlier items. In addition, respondents with more  correctly answered items (that is, higher sum scores) have higher $\theta_k$ estimates. Lastly, the distribution of the posterior samples for $\alpha_i$ is similar across all items.

\begin{figure}[htpb]
    \centering
    \begin{subfigure}[b]{0.4\textwidth}
        \centering
        \includegraphics[width=\textwidth]{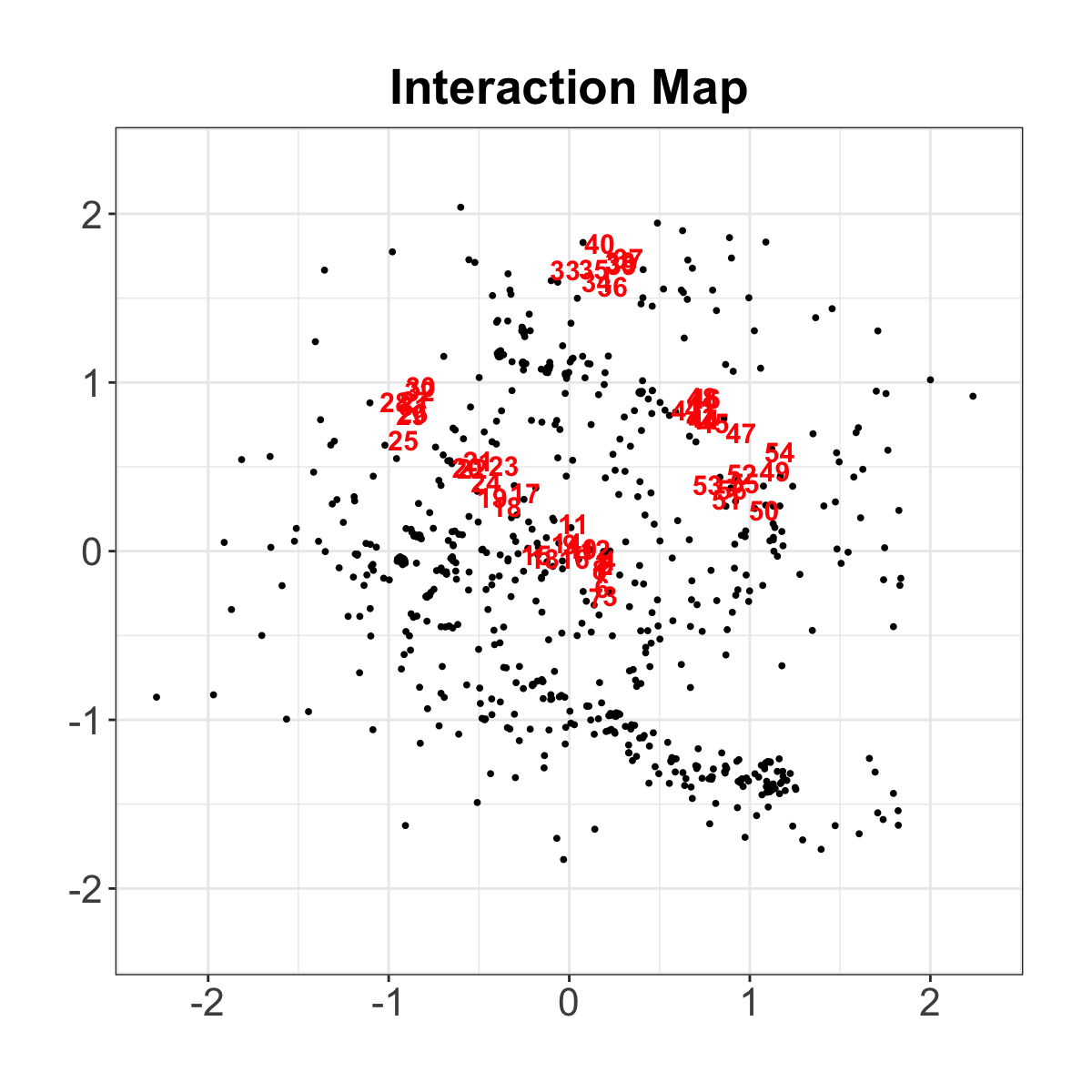}
        \caption{Original} \label{fig:2pl_latent_mapa}
    \end{subfigure}
    \hspace{0.05\textwidth}
    \begin{subfigure}[b]{0.4\textwidth}
        \centering
        \includegraphics[width=\textwidth]{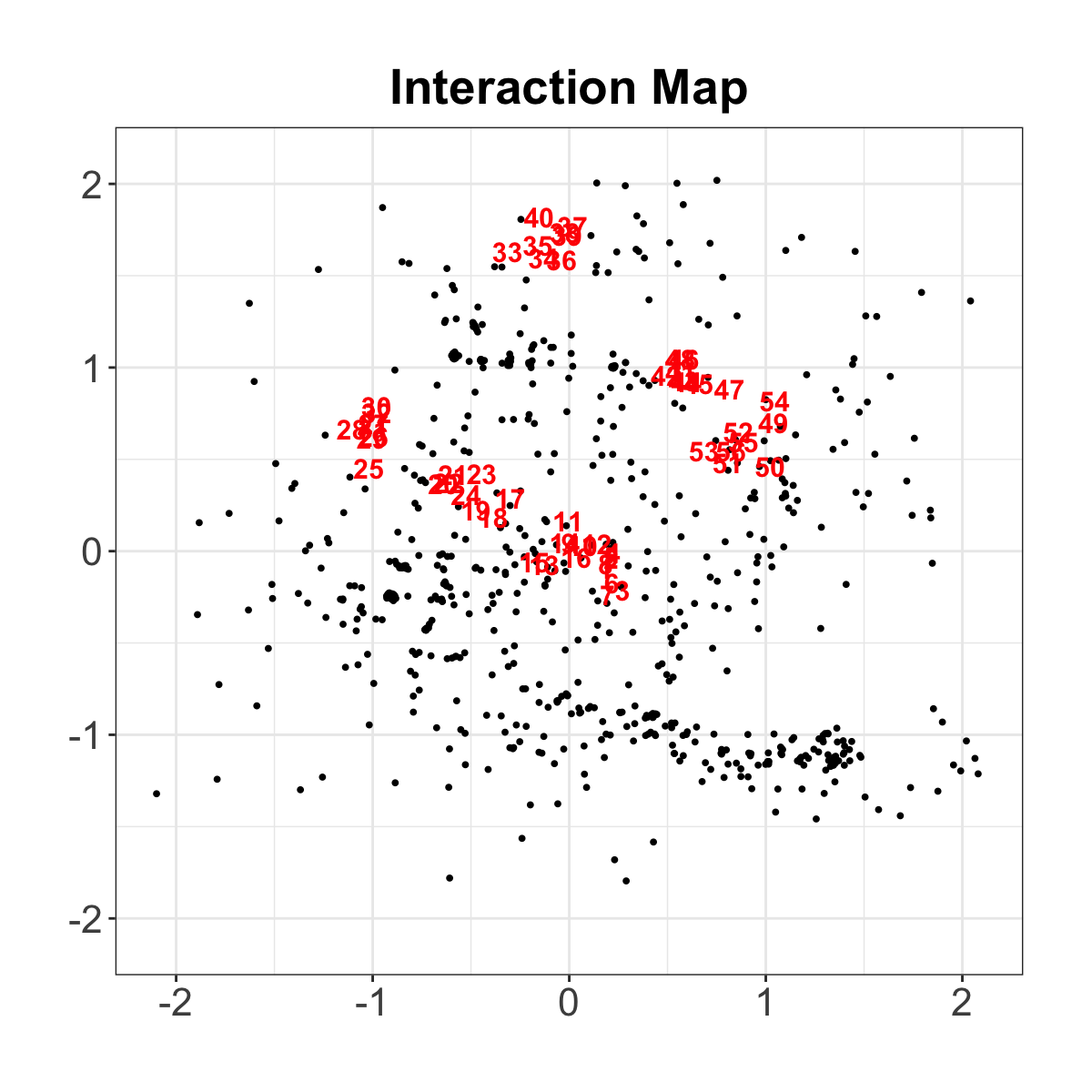}
        \caption{Oblimin rotation} \label{fig:2pl_latent_mapb}
    \end{subfigure}
    \caption{
    The interaction map based on the 2PL LSIRM fitted to the TDRI dataset. (a) Visualization of the interaction map based on the 2PL LSIRM. (b) A rotated interaction map using oblimin rotation. Red numbers and black dots represent the latent positions for items and respondents, respectively, in both plots. 
    }
    \label{fig:2pl_latent_map}
\end{figure}

The \code{plot()} function can be used to create a visualization of the interaction map based on the 2PL LSIRM by setting \code{option} argument as \code{"interaction"}. Figures \ref{fig:2pl_latent_mapa} and \ref{fig:2pl_latent_mapb} show the original and rotated interaction maps, respectively. In the interaction maps based on the 1PL and 2PL LSIRM, we notice that although the clustering of items is similar, the distances between item groups appear smaller in the interaction map of the 2PL LSIRM. This difference arises because the 2PL LSRIM takes item discrimination ($\alpha$) into account, which explains some degree of item-by-person interactions. The interaction map with the oblimin rotation (b) shows a slight clockwise rotation compared to the original interaction map (a), similarly to the 1PL LSIRM's case. 
%The interaction map based on the 2PL LSIRM appears similar to the interaction map from the 1PL LSIRM, but there are some differences. 
%, the dislatent positions of the items appear closer to the latent positions of the respondents compared to the 1PL LSIRM.
% that are not captured by the 1PL LSIRM.
%, aligned with the axis direction. 

% \textcolor{red}{[[explain that the results between the original and rotated plots are similar]]}

% \textcolor{red}{[[briefly explain what these differences are.]]} 

\begin{example}
R > plot(lsirm_result, option = "interaction", chain.idx = 1)
R > plot(lsirm_result, option = "interaction", rotation = TRUE, chain.idx = 1)
\end{example}

%Similar to the 1PL LSIRM, 
The \code{plot()} function visualizes the cluster of the latent positions of items by using spectral clustering and the Neyman-Scott process modeling approach, achieved by setting the \code{cluster} option to \code{spectral} and \code{neyman}, respectively.
\begin{example}
R > plot(lsirm_result, cluster = "spectral", chain.idx = 1)

Clustering result (Spectral Clustering): 
  group                                 item 
      A         49, 50, 51, 52, 53, 54, 55, 56 
      B         25, 26, 27, 28, 29, 30, 31, 32, 33, 34,
                35, 36, 37, 38, 39, 40 
      C         41, 42, 43, 44, 45, 46, 47, 48   
      D         1, 2, 3, 4, 5, 6, 7, 8, 9, 10,
                11, 12, 13, 14, 15, 16  
      E         17, 18, 19, 20, 21, 22, 23, 24 

 R > plot(lsirm_result, cluster = "neyman", chain.idx = 1)
 
|==================================================| 100%
Clustering result (Neyman-Scott process): 
  group                                 item 
      A         33, 34, 35, 36, 37, 38, 39, 40 
      B         49, 50, 51, 52, 53, 54, 55, 56 
      C         41, 42, 43, 44, 45, 46, 47, 48 
      D         17, 18, 19, 20, 21, 22, 23, 24 
      E         25, 26, 27, 28, 29, 30, 31, 32 
      F         1, 2, 3, 4, 5, 6, 7, 8 
      G         9, 10, 11, 12, 13, 14, 15, 16  
\end{example}
\begin{figure}[htb]
    \centering
    \begin{subfigure}[b]{0.4\textwidth}
        \centering
        \includegraphics[width=\textwidth]{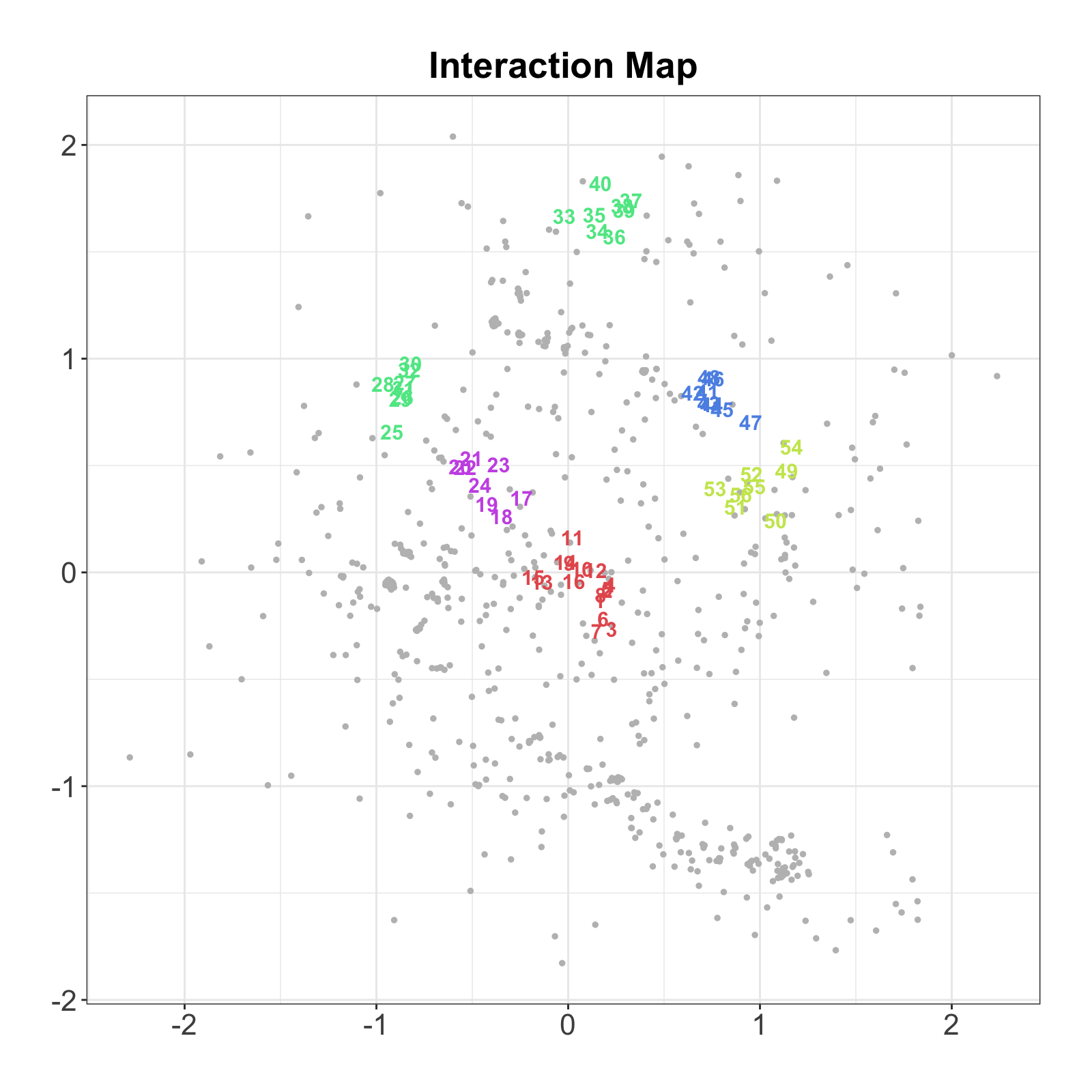} 
        \caption{Spectral Clustering} \label{fig:2pl_latent_map_clustera}
    \end{subfigure}
    \hspace{0.05\textwidth}
    \begin{subfigure}[b]{0.4\textwidth}
        \centering
        \includegraphics[width=\textwidth]{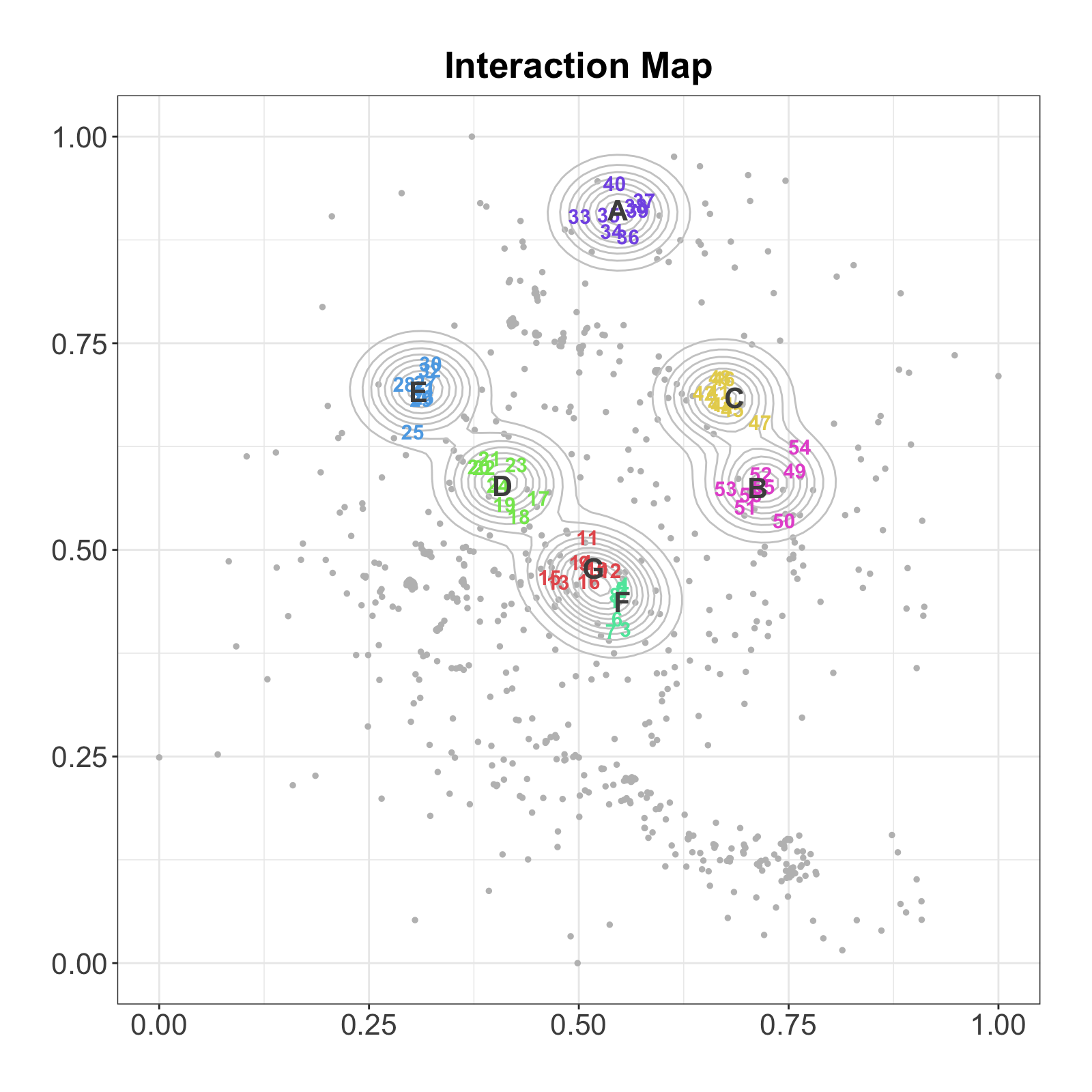}
        \caption{Neyman-Scott Process Model} \label{fig:2pl_latent_map_clusterb}
    \end{subfigure}
    \caption{
    The interaction map with the item clustering results based on the 2PL LSIRM fitted to the TDRI dataset, using (a) spectral clustering and (b) the Neyman-Scott process modeling approach. In both plots, the gray dots indicate respondents, where numbers in colors indicate items with different cluster memberships. The Neyman-Scott process approach additionally displays the center of the cluster (alphabets) and a contour for each cluster.
    }
    \label{fig:2pl_latent_map_cluster}
\end{figure}
Figure \ref{fig:2pl_latent_map_clustera} and  \ref{fig:2pl_latent_map_clusterb} display the result of spectral clustering and the Neyman-Scott process approach, respectively. The interpretation of gray dots, numbers, alphabets, and contours is the same as the 1PL LSIRM case. Spectral clustering resulted in 5 clusters, whereas the Neyman-Scott process approach resulted in 7 clusters. That is, the Neyman-Scott process approach identified more item groups than the spectral clustering. The overall clustering results are similar to the 1PL LSIRM case. 
%When compared  the 1PL LSIRM results, the cluster results were similar. Additionally, the Neyman-Scott process model produced more detailed cluster results, similar to the 1PL LSIRM.

%\textcolor{red}{[[mj: explain the results briefly]]}

% such as \code{fixed\_gamma}, \code{spikenslab}, and \code{missing\_data} 
The flexible modeling options discussed with the 1PL LSIRM can be applied to the 2PL LSIRM. That is, \code{fixed\_gamma} fixes the distance weight $\gamma$ to 1 and \code{spikenslab} assigns a spike and slab prior to $\gamma$. Additionally, the two missing data options, \code{missing\_data = "mcar"} and \code{missing\_data = "mar"}, are available for the 2PL LSIRM.

\begin{example}
R > lsirm_result <- lsirm(data ~ lsirm2pl(fixed_gamma = TRUE))
R > lsirm_result <- lsirm(data ~ lsirm2pl(spikenslab = TRUE))
R > lsirm_result <- lsirm(data ~ lsirm2pl(missing_data = "mcar"))
R > lsirm_result <- lsirm(data ~ lsirm2pl(missing_data = "mar"))
\end{example}

\section{LSIRM For Continuous Item Responses Data}\label{conti}

We consider an extension of the LSIRM for continuous item response data (LSIRM-continuous), which is done by using an appropriate link function following the generalized linear model framework \citep{mccullagh2019generalized}. 

\subsection{Statistical framework}

Consider the continuous item response data consisting of the $N \times P$ matrix $\boldsymbol{Y} = \left\{ y_{k,i} \right\} \in \mathbb{R}^{N\times P}$. By choosing the identity link function for continuous item responses, the 1PL LSIRM-continuous is given as follows:
\begin{align*}
        y_{k,i} \mid \boldsymbol{\theta}, \boldsymbol{\beta}, \boldsymbol{\gamma},\boldsymbol{Z}, \boldsymbol{W},\sigma_{\epsilon}^2 = \theta_{k}+\beta_{i}-\gamma d (\boldsymbol{z_{k}},\boldsymbol{w_{i}}) + \epsilon_{k,i}, \; \; \theta_{k} \sim \mbox{N}(0, \sigma^2), \; \; 
        \epsilon_{k,i} \sim \mbox{N} (0, \sigma_{\epsilon}^2).
\end{align*}
where $\theta_k$ and $\beta_i$ are the main effects of the respondent $k$ and the item $i$, respectively. $\boldsymbol{z_k}$ and $\boldsymbol{w_i}$ are the latent positions of the respondent $k$ and the item $i$, respectively. An additional error term $\epsilon_{k,i} \sim N(0, \sigma_{\epsilon}^2)$ explains residuals unexplained by the 1PL LSIRM-continuous. A shorter distance between $\boldsymbol{w_i}$ and $\boldsymbol{z_k}$ indicates that the respondent $k$ is likely to give a higher response value to item $i$ given the main effects of the respondent and the item.
%In other words, the model specifies that $y_{k,i}$ follows a normal distribution with mean $\theta_k +\beta_i - \gamma d (\boldsymbol{z_{k}}, \boldsymbol{w_{i}})$ and variance $\sigma_{\epsilon}^2$. 

It is straightforward to extend the 1PL LSIRM-continuous to the 2PL version by adding item discrimination (or slope) parameters $\alpha_i$. The 2PL LSIRM-continuous for $y_{k,i}$ is given as follows:
\begin{align*}
        y_{k,i} \mid \boldsymbol{\theta},\boldsymbol{\alpha}, \boldsymbol{\beta}, \boldsymbol{\gamma},\boldsymbol{Z}, \boldsymbol{W},\sigma_{\epsilon}^2 = \alpha_{i}\theta_{k}+\beta_{i}-\gamma d (\boldsymbol{z_{k}},\boldsymbol{w_{i}}) + \epsilon_{k,i}, \; \; \theta_{k} \sim \mbox{N}(0, \sigma^2), \; \; 
        \epsilon_{k,i} \sim \mbox{N} (0, \sigma_{\epsilon}^2).
\end{align*}
The interpretations of the model parameters in the 2PL LSIRM-continuous are similar to the case of the 1PL LSIRM-continuous and the 2PL LSIRM. For model identifiability, one of the item slopes is fixed to  1, e.g., $\alpha_{1} =1$. 

The likelihood function of the 1PL LSIRM-continuous is given as 
\begin{align*}
\mathbf{L}\left(\boldsymbol{Y=y|\theta,\beta},\gamma,\boldsymbol{Z,W},\sigma_{\epsilon}^2\right)
& =\prod_{k=1}^{N}\prod_{i=1}^{P}\mathbf{L}(Y_{k, i}=y_{k, i}|\theta_{k}, \beta_{i},\gamma,\boldsymbol{z_{k},w_{i}},\sigma_{\epsilon}^2), \\
& = \prod_{k=1}^{N}\prod_{i=1}^{P} N(y_{k, i,};\theta_k +\beta_i - \gamma || {\boldsymbol{z_k}} - {\boldsymbol{w_i}} ||, \sigma_{\epsilon}^2),
\end{align*}
and the likelihood function of the 2PL LSIRM-continuous is given as
\begin{align*}
\mathbf{L}\left(\boldsymbol{Y=y|\theta,\alpha,\beta},\gamma,\boldsymbol{Z,W},\sigma_{\epsilon}^2\right)
& = \prod_{k=1}^{N}\prod_{i=1}^{P} N(y_{j, i};\alpha_i\theta_k +\beta_i - \gamma|| {\boldsymbol{z_k}} - {\boldsymbol{w_i}} ||, \sigma_{\epsilon}^2).
\end{align*}

\vspace*{-0.75cm}

% % 2PL normal posteior
% \begin{align*}
% \pi\left(\boldsymbol{\theta, \beta, \alpha}, \gamma, \boldsymbol{Z, W}, \sigma_{\epsilon}^2 | \boldsymbol{Y=y} \right) &\propto \prod_{k=1}^{N}\prod_{i=1}^{P} N(y_{j, i};\alpha_i\theta_k +\beta_i - \gamma|| {\boldsymbol{z_k}} - {\boldsymbol{w_i}} ||, \sigma_{\epsilon}^2) \\
% &\times \prod_{k=1}^N \pi(\theta_k) \times \prod_{i=1}^P \pi(\alpha_i) \times \prod_{i=1}^P \pi(\beta_i) \times \pi(\gamma) \times \prod_{k=1}^N \pi(\boldsymbol{z_{k}}) \times \prod_{i=1}^P \pi(\boldsymbol{w_{i}}) \times \pi(\sigma_{\epsilon}^2)
% \end{align*}

\subsection{Parameter Estimation}
Following is the posterior distribution of LSRIM-continuous.
\begin{align*}
\pi\left(\boldsymbol{\theta, \beta}, \gamma, \boldsymbol{Z, W}, \sigma_{\epsilon}^2 | \boldsymbol{Y=y} \right) &\propto \prod_{k=1}^{N}\prod_{i=1}^{P}N(y_{k, i,};\theta_k +\beta_i - \gamma || {\boldsymbol{z_k}} - {\boldsymbol{w_i}} ||, \sigma_{\epsilon}^2) \\
&\times \prod_{k=1}^N \pi(\theta_k) \times \prod_{i=1}^P \pi(\beta_i) \times \pi(\gamma) \times \prod_{k=1}^N \pi(\boldsymbol{z_{k}}) \times \prod_{i=1}^P \pi(\boldsymbol{w_{i}}) \times \pi(\sigma_{\epsilon}^2)
\end{align*}
The priors of the model parameters in the LSIRM-continuous are set the same as the priors for the LSIRM. For the error term $\sigma_{\epsilon}^2$, we set the prior as follows:
\begin{align*}
\sigma_{\epsilon}^{2}|a_{\sigma_{\epsilon}},b_{\sigma_{\epsilon}} & \sim\text{Inv-Gamma}(a_{\sigma_{\epsilon}},b_{\sigma_{\epsilon}}),\ a_{\sigma_{\epsilon}}>0,\ b_{\sigma_{\epsilon}}>0.
\end{align*}
The conditional posterior distributions of the LSIRM-continuous are similar to the LSIRM, which are given in the Github site. The jumping rule defaults remain the same as in the binary case (shown in the Github site). 

\subsection{An Illustrated Example}

We demonstrate the application of the LSIRM-continuous using the Big Five Personality Test (FPT) dataset \citep{Goldberg1992}, which is included in the package. Data were collected from 1,015,342 respondents using an interactive online personality test from 2016 to 2018. The ``Big-Five Factor Markers'' from the international personality item pool were used in the test, which consists of 50 questions with response categories 1 = disagree, 3 = neutral, and 5 = agree based on a five-point Likert scale. Negatively worded items were reverse-coded. For illustration purposes, we randomly selected a sample of 3,000 respondents from the original data. We treated the ordinal item responses as continuous data, which is a common practice in applied research and is generally considered acceptable for other models, such as factor analysis. 
 %\footnote{The dataset can be downloaded for reproducible purposes from the following link: \url{https://www.kaggle.com/tunguz/big-five-personality-test}}. 

The main fitting function for running the 1PL and 2PL LSIRM-continuous is identical to the 1PL and 2PL LSIRM for binary data.  The function automatically identifies the data type and applies the appropriate models. Following is the code for data pre-processing and 1PL LSIRM-continuous fitting.

\begin{example}
R > data <- lsirm12pl::BFPT
R > data[(data==0)|(data==6)] = NA
R > reverse <- c(2, 4, 6, 8, 10, 11, 13, 15, 16, 17,
                 18, 19, 20, 21, 23, 25, 27, 32, 34,
                 36, 42, 44, 46)
R > data[, reverse] <- 6 - data[, reverse]
R > data <- data[complete.cases(data),]
R > head(data)
    EXT1 EXT2 EXT3 ... OPN8 OPN9 OPN10
1    2    3    2   ...  2    4    4
2    1    3    3   ...  4    3    4
3    4    3    3   ...  2    4    4
5    1    4    3   ...  3    5    3
6    1    3    2   ...  3    5    3
8    3    5    4   ...  4    3    4

R > lsirm_result <- lsirm(data ~ lsirm1pl(niter = 25000, nburn = 5000, nthin = 10,
                                      jump_beta = 0.08, jump_theta = 0.3,
                                      jump_gamma  = 1.0, 
                                      chains = 4, multicore = 2, seed = 2025))
\end{example}

To ensure convergence of the parameters, a different number of iteration and jumping rules of parameters are applied in this example. The estimated results for each chain are returned in the list containing estimated information on $\boldsymbol{\theta,\ \beta},\ \gamma,\ \boldsymbol{Z,\ W},\ {\sigma}$, and $\sigma_\epsilon$. The functions \code{summary()}, \code{diagnostics()}, and \code{gof()} described for the 1PL LSIRM can be applied similarly to obtain a summary, diagnosis, and goodness-of-fit results, respectively.

\begin{example}
R > diagnostic(lsirm_result, draw.item = list(beta = c("AGR1")))
\end{example}

\begin{figure}[htb]
    \centering
    \includegraphics[width=.8\textwidth]{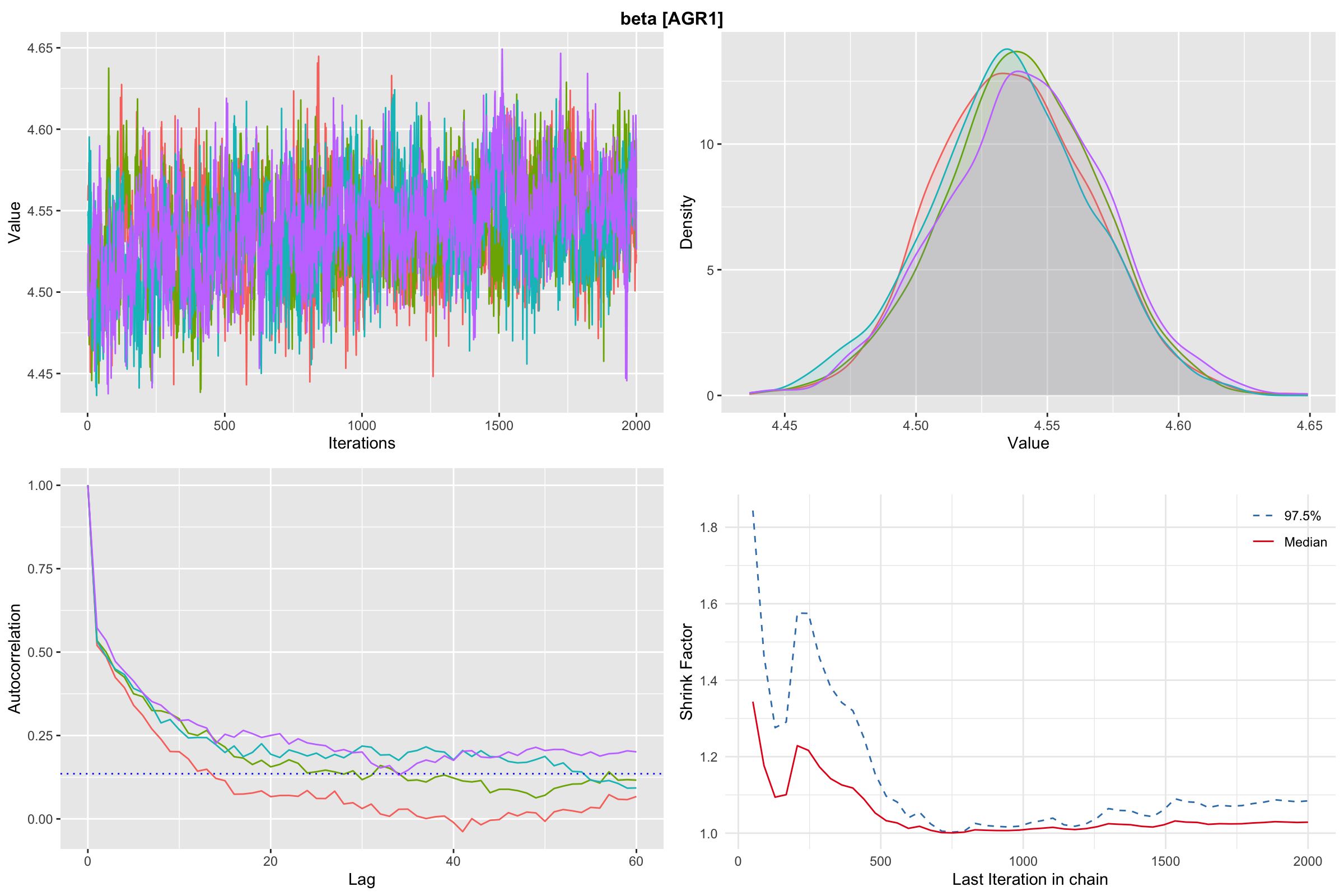}
    \caption{ 
    Diagnostics of $\beta_{AGE_1}$ using the \code{diagnostic()} function on the results of the 1PL LSIRM-continuous fitted to the FPT dataset. The top left is the trace plot, the top right is the posterior density plot, the bottom left is the autocorrelation plot, and the bottom right is the Gelman-Rubin-Brooks plot. Different colors represent different MCMC chains.
    }
    \label{fig:diagnosis1-cont}
\end{figure}

Figure \ref{fig:diagnosis1-cont} displays the diagnostic results for $\beta_{AGE_1}$ obtained using the \code{diagnostic()} function. The results suggest the convergence of $\beta_{AGE_1}$ is achieved for this model.

\begin{figure}[htb]
    \centering
    \includegraphics[width=0.4\textwidth]{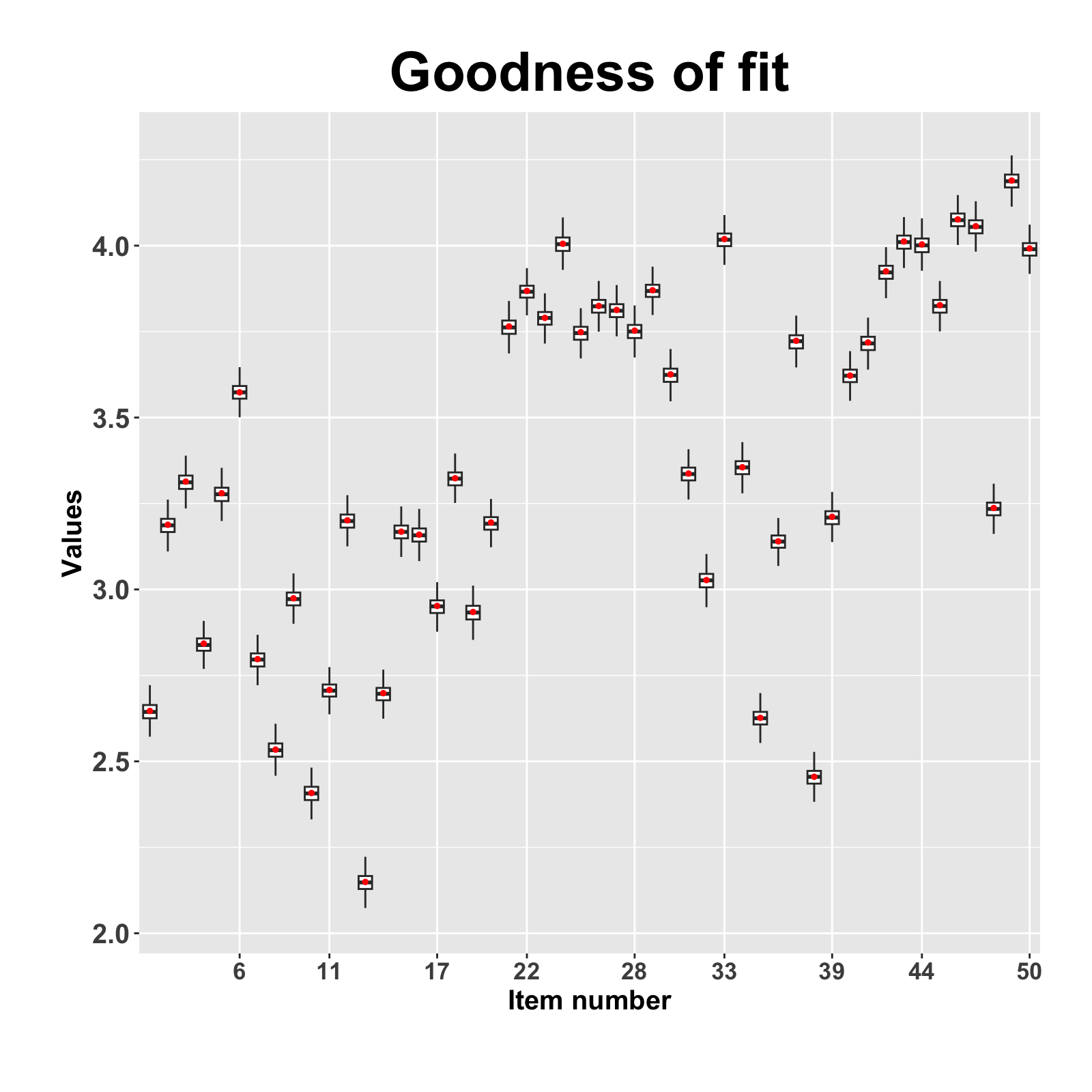}
    \caption{
    Goodness of fit of the 1PL LSIRM-continuous for the BFPT dataset using the \code{gof()} function. 
    The box plots of the average posterior predicted response value for items against the average observed response value for each item indicated by red dots.
    }
    \label{fig:gof3}
\end{figure}

Figure \ref{fig:gof3} shows the result of the goodness of fit assessment for the continuous example data. Unlike binary data, ROC is not available for continuous data, so the results of the \code{gof()} function include only the boxplots of the average predicted response values (item-wise means) against the observed item-wise means (marked with red dots). In this example, the red dots are located close to the midlines of the boxplots, implying a satisfactory model fit.

%\textcolor{red}{[[explain what red dots are. some red dots are not close to the midlines/ Solve! (Changes the posterior predicted response value calculation method used when drawing a boxplot.)]]}

\begin{example}
R > gof(lsirm_result, chain.idx = 1)
\end{example}

Same as before, the \code{plot()} function can be used to summarize the parameter estimate of $\beta_i$ and $\theta_k$. % by setting \code{option} argument as \code{beta} and \code{theta}, respectively. 

\begin{example}
R > plot(lsirm_result, option = "beta", chain.idx = 1)
R > plot(lsirm_result, option = "theta", chain.idx = 1) 
\end{example}

\begin{figure}[ht]
     \centering
     \begin{subfigure}[b]{0.35\textwidth}
         \centering
         \includegraphics[width=\textwidth]{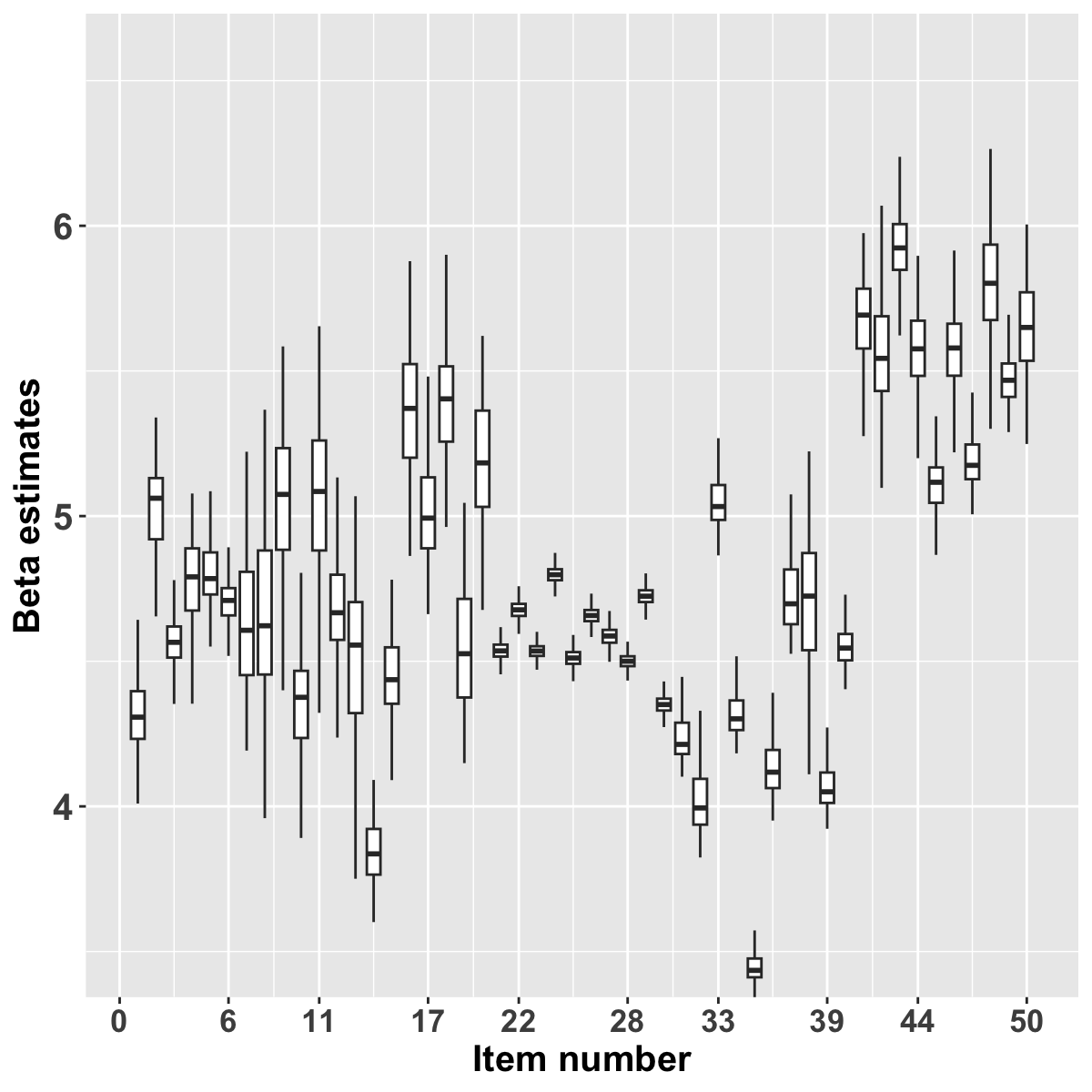}
         \caption{$\beta_i$} \label{fig:latent_param_1pl_normala}
     \end{subfigure}
     \hspace{.3cm}
     \begin{subfigure}[b]{0.35\textwidth}
         \centering
         \includegraphics[width=\textwidth]{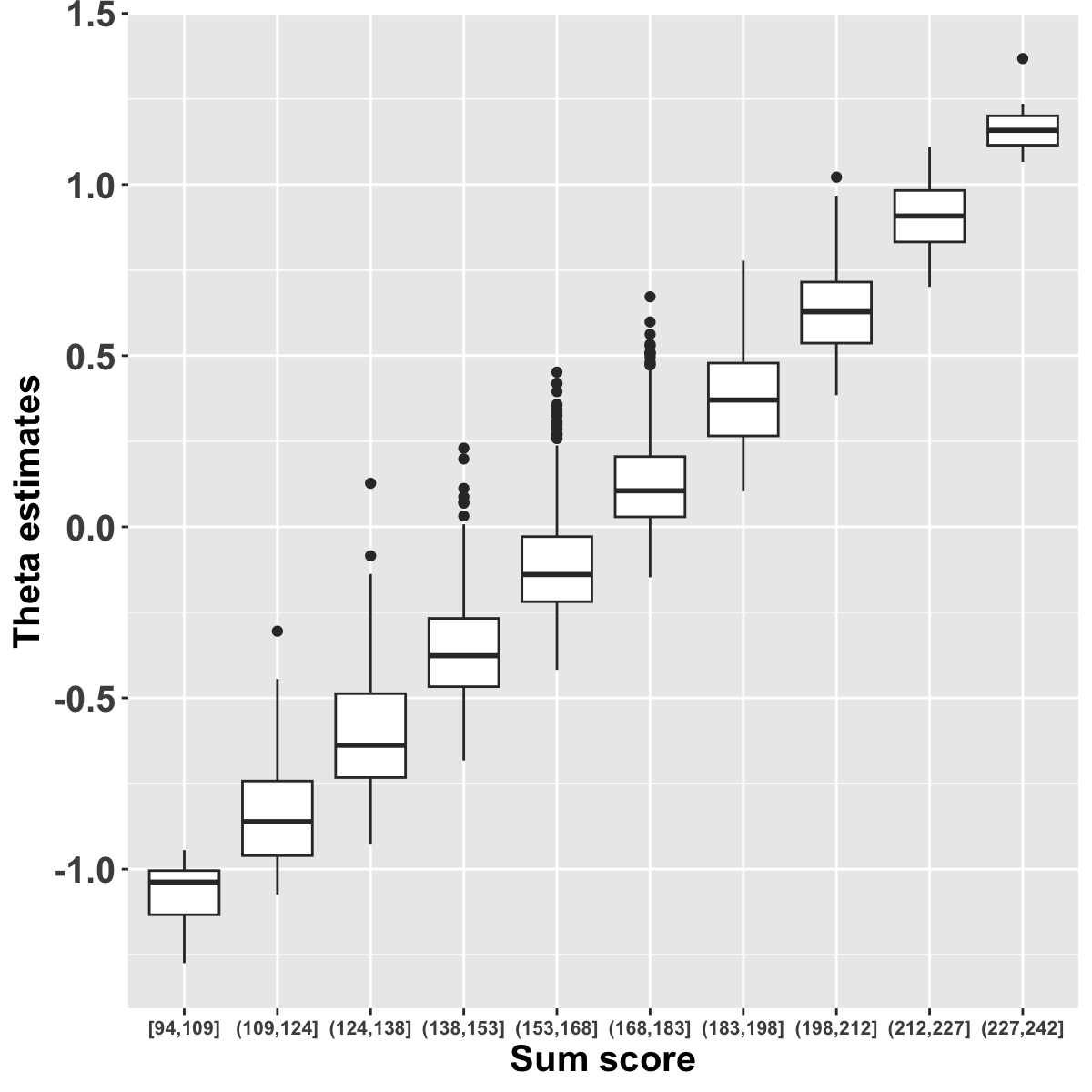}
         \caption{$\theta_k$} \label{fig:latent_param_1pl_normalb}
     \end{subfigure}
     \caption{
     Visual summaries of $\beta_i$ and $\theta_k$ using the \code{plot()} function on the 1PL LSIRM-continuous fitted to the FPT dataset. (a) Boxplots of the posterior samples for $\beta_i$. (b) Boxplots of the point estimates of $\theta_k$ as a function of the total sum scores of the responses binned into 10 groups
     }
    \label{fig:latent_param_1pl_normal}
\end{figure}

Figure \ref{fig:latent_param_1pl_normala} shows the boxplots of the posterior samples for $\beta_i$. The parameter estimate of $\beta_{35}$ is the smallest , while the estimates for $\beta_{41}$ to $\beta_{50}$ are relatively high. Figure \ref{fig:latent_param_1pl_normalb} shows the boxplots of the point estimates of $\theta_k$ as a function of the sum scores of the observed responses binned into 10 groups to prevent overlap and improve readability on the x-axis. As expected, higher sum scores are aligned with higher $\theta_k$ values (indicating socially desirable characteristics). 
%This means that the 35th question has the highest threshold, meaning  whereas the 41 to 50 questions are easier and more likely to be answered correctly by most individuals.
% \textcolor{red}{mj: mention and explain about "binned into 10 groups"}

\begin{figure}[ht]
     \centering
     \begin{subfigure}[b]{0.4\textwidth}
         \centering
         \includegraphics[width=\textwidth]{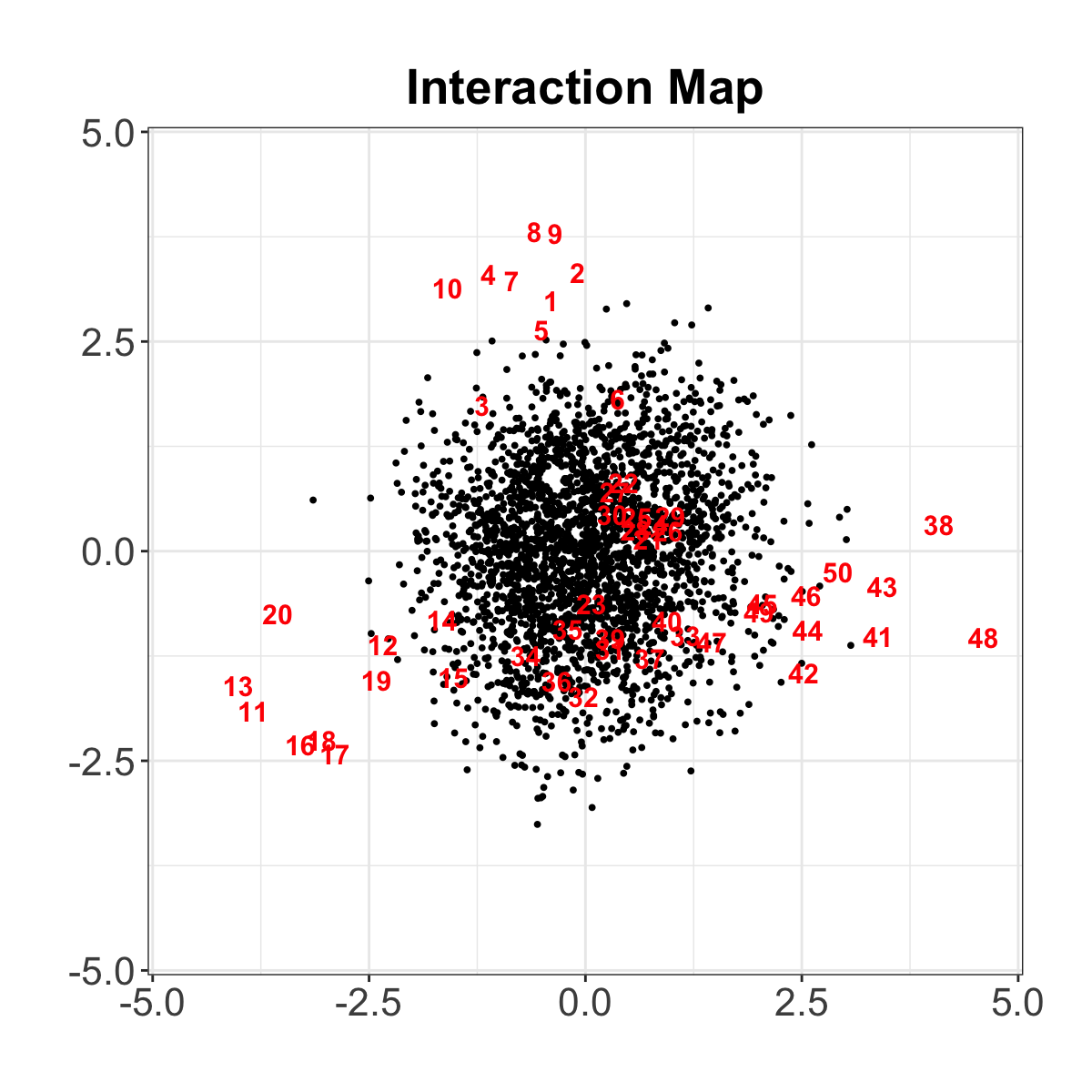}
         \caption{Original} \label{fig:1pl_normal_latent_mapa}
     \end{subfigure}
     \hspace{.3cm}
     \begin{subfigure}[b]{0.4\textwidth}
         \centering
         \includegraphics[width=\textwidth]{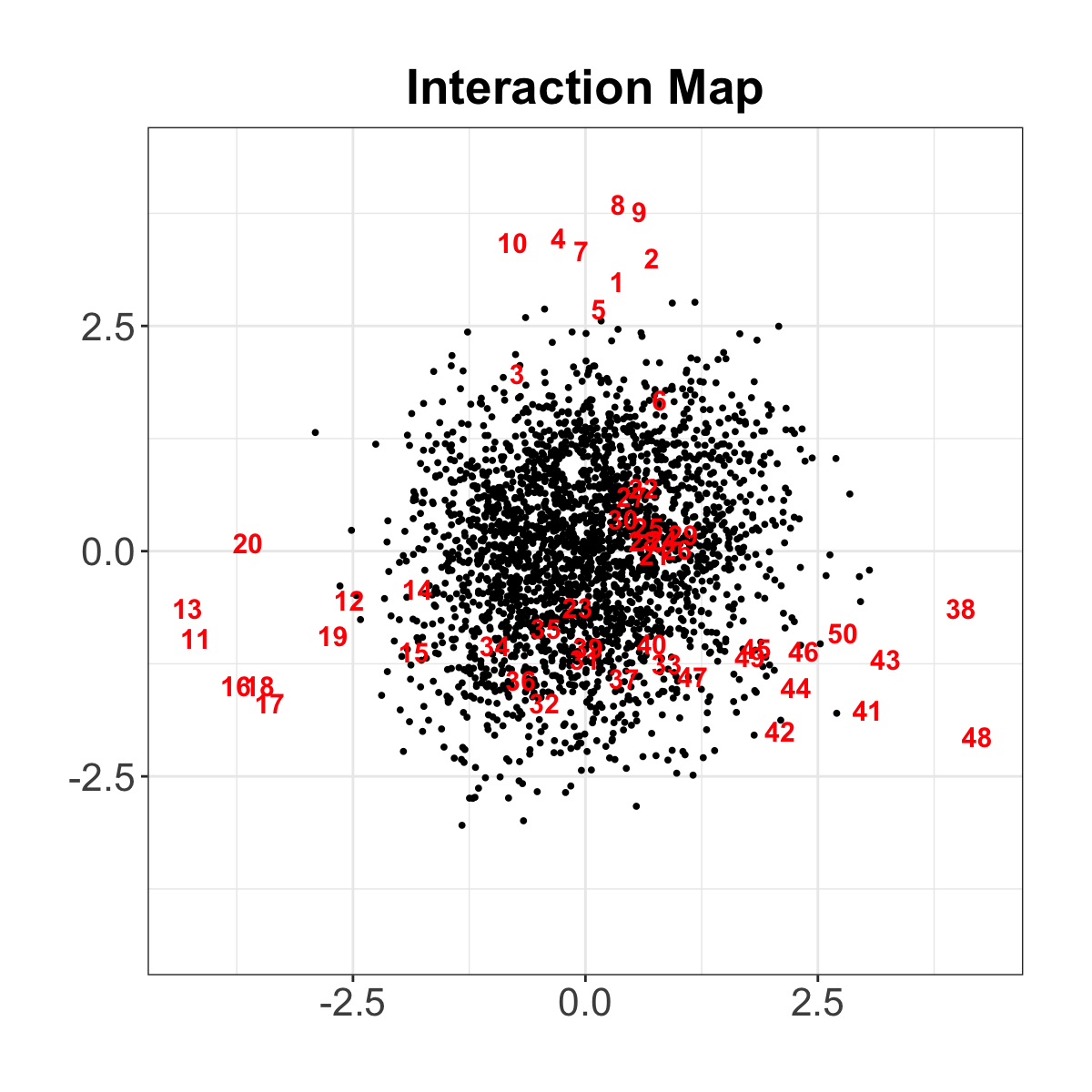} 
         \caption{Oblimin rotation} \label{fig:1pl_normal_latent_mapb}
     \end{subfigure}
    \caption{
    The interaction map based on the 1PL LSIRM-continuous fitted to the FPT dataset. (a) Visualization of the interaction map based on the 1PL LSIRM-continuous. (b) A rotated interaction map of the 1PL LSIRM-continuous using oblimin rotation. Red numbers and black dots represent the latent positions for items and respondents, respectively.
    }
    \label{fig:1pl_normal_latent_map}
\end{figure}

Figure \ref{fig:1pl_normal_latent_mapa} illustrates the interaction map derived from the 1PL LSIRM-continuous. In the interaction map, the latent positions of the respondents are positioned  around the center, while the items are scattered in a triangular pattern showing roughly three clusters (in the north, west, and east) in addition to the cluster around the center. The original interaction map is slightly rotated counterclockwise in the rotated interaction map, so that the the items are more closely placed to the two coordinates. 
%When the distance between the latent position of the respondent and the item is large, the respondent is likely to give a low response value to the item. 
%The original and rotated interaction map with oblimin rotation are similar. This is because the items in original interaction map was already placed close to the two coordinates. 

%\textcolor{red}{[[explain the results more, for instance about the item clusters. also the differences between the original and rotated plots]]}

\begin{example}
R > plot(lsirm_result, chain.idx = 1)
R > plot(lsirm_result, rotation = TRUE, chain.idx = 1)
\end{example}

\begin{figure}[ht]
     \centering
     \begin{subfigure}[b]{0.4\textwidth}
         \centering
         \includegraphics[width=\textwidth]{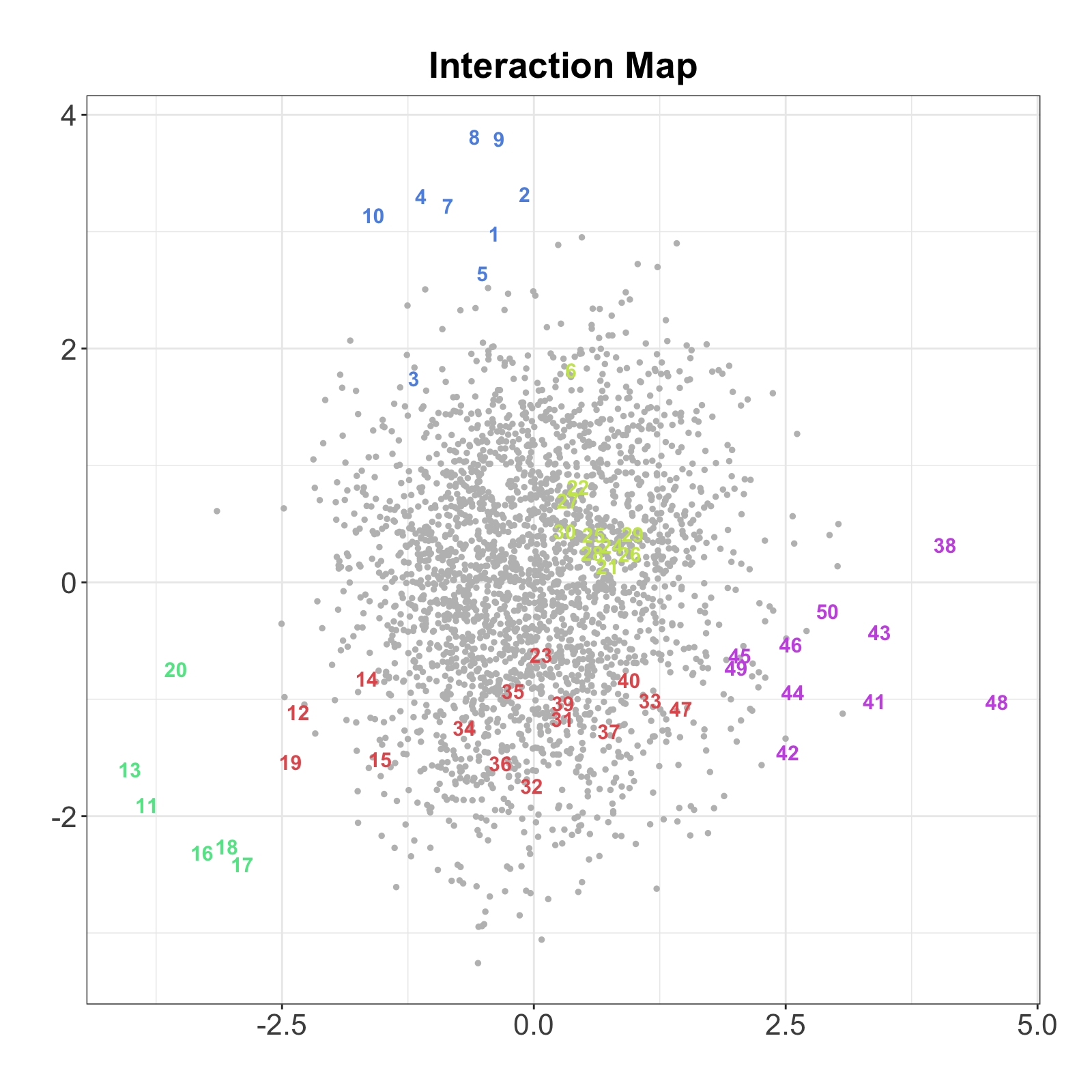}
         \caption{Spectral Clustering} \label{fig:1pl_cont_latent_map_clustera}
     \end{subfigure}
     \hspace{.3cm}
     \begin{subfigure}[b]{0.4\textwidth}
         \centering
         \includegraphics[width=\textwidth]{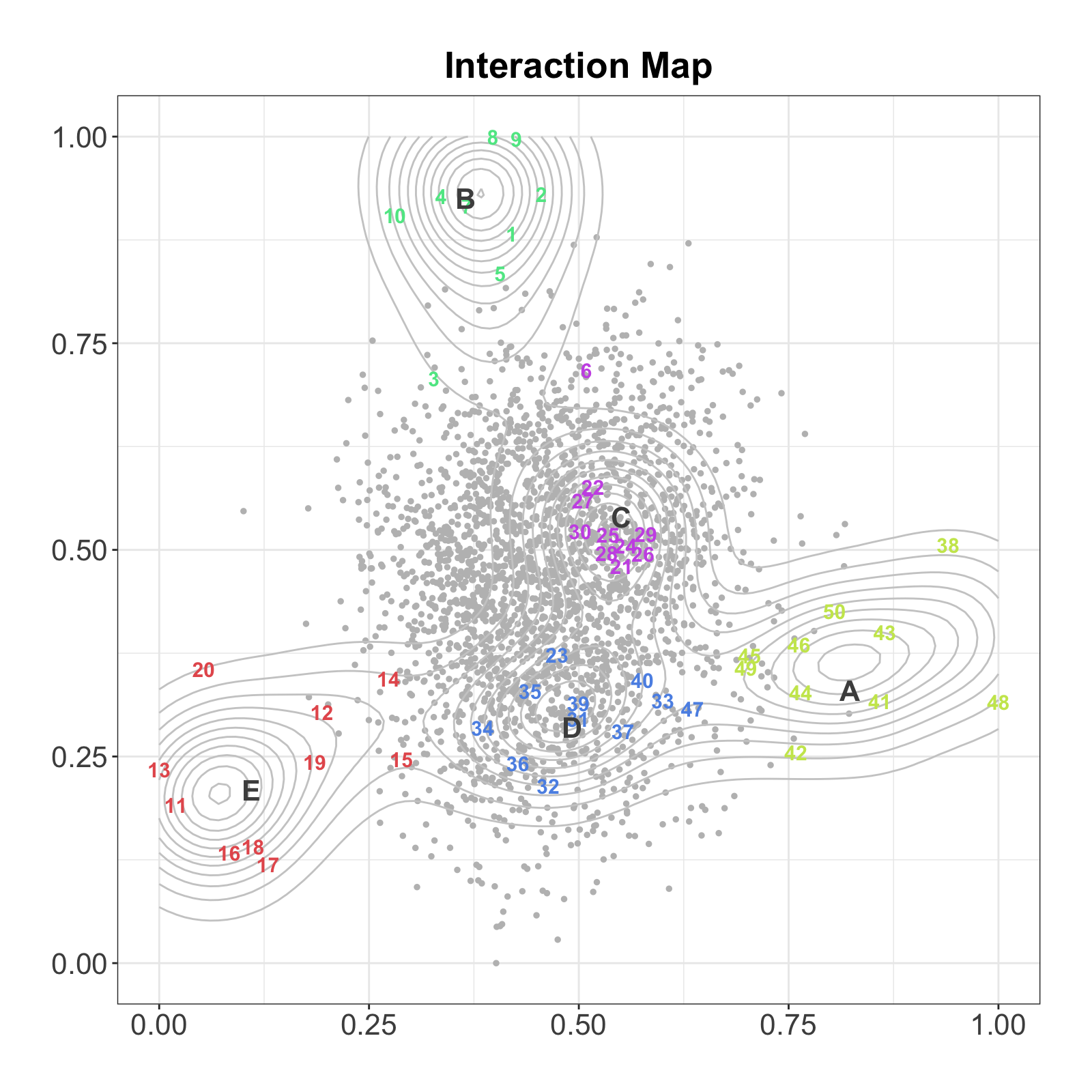}
         \caption{Neyman-Scott Process Model} \label{fig:1pl_cont_latent_map_clusterb}
     \end{subfigure}
    \caption{
    The interaction map with item clustering results based on the 1PL LSIRM model-continuous fitted to the FPT dataset, using (a) spectral clustering and (b) Neyman-Scott process approach. In both plots, the gray dots indicate respondents, where numbers in colors indicate items with different cluster memberships. The Neyman-Scott process approach additionally displays the center of the cluster (alphabets) and a contour for each cluster. 
    }
    \label{fig:1pl_cont_latent_map_cluster}
\end{figure}

Figure \ref{fig:1pl_cont_latent_map_clustera} and Figure \ref{fig:1pl_cont_latent_map_clusterb} depict the result of spectral clustering and the Neyman-Scott process modeling approach for the BFPT example dataset, respectively. Gray dots, numbers with colors, alphabets, and contours have the same interpretations as the clustering results presented earlier with the binary LSIRM models. In Figure \ref{fig:1pl_cont_latent_map_clusterb}, the items in cluster C, highlighted in purple and located near the center of the interaction map, are closer to the latent positions of most people compared to other items. This implies that these items are more likely to receive higher response values. Conversely, items in clusters that are positioned farther from the center, such as clusters A, B, and E, are  distant from many (but different groups of) respondents, indicating they are likely to receive lower response values by those who are far away  from the corresponding item clusters.

\begin{example}
R > plot(lsirm_result, cluster = "spectral", chain.idx = 1)

Clustering result (Spectral Clustering): 
  group                                       item 
      A         11, 13, 16, 17, 18, 20  
      B         38, 41, 42, 43, 44, 45, 46, 48, 49, 50 
      C         1, 2, 3, 4, 5, 7, 8, 9, 10 
      D         6, 21, 22, 24, 25, 26, 27, 28, 29, 30 
      E         12, 14, 15, 19, 23, 31, 32, 33, 34, 35,   
                36, 37, 39, 40, 47 

R > plot(lsirm_result, cluster = "neyman", chain.idx = 1)

|==================================================| 100%
Clustering result (Neyman-Scott process): 
  group                                       item 
      A         38, 41, 42, 43, 44, 45, 46, 48, 49, 50  
      B         1, 2, 3, 4, 5, 7, 8, 9, 10 
      C         6, 21, 22, 24, 25, 26, 27, 28, 29, 30 
      D         23, 31, 32, 33, 34, 35, 36, 37, 39, 40, 47 
      E         11, 12, 13, 14, 15, 16, 17, 18, 19, 20  
\end{example}

The flexible modeling options discussed with the binary LSIRM can be applied to the functions for the LSIRM-continuous. 

\begin{example}
R > lsirm_result <- lsirm(data ~ lsirm1pl(fixed_gamma = TRUE))
R > lsirm_result <- lsirm(data ~ lsirm1pl(spikenslab = TRUE))
R > lsirm_result <- lsirm(data ~ lsirm1pl(missing_data = "mcar"))
R > lsirm_result <- lsirm(data ~ lsirm1pl(missing_data = "mar"))
\end{example}

\section{Conclusion}\label{conclusion}

In this paper, we introduced the R package \CRANpkg{lsirm12pl} for estimating the LSIRM \citep{jeon2021mapping} and its extensions. The LSIRM is a powerful extension of conventional IRT models that allow for the estimation and visualization of potential interactions between respondents and items in an interaction map, a low dimensional Euclidean space. The original LSIRM framework was proposed for binary item response data based on the 1PL IRT base model. To broaden its applicability, we extended the original LSIRM to cover different response types and model specifications. Further, we added several useful options, e.g.,  for handling missing data, item clustering, and model assessment. 

% Second revision
It is worth noting that one may observe some  variability in clustering results across multiple chains. % as a natural result of unsupervised clustering methods and the Bayesian framework. 
Such variability stems from at least three factors: First, clustering is an unsupervised method, meaning that results are not uniquely determined and may naturally vary across Markov chains. Second, we apply the Neyman-Scott Process clustering  to the posterior distributions of the LSIRM model, and the clustering outcomes are therefore directly influenced by the estimated LSIRM results. Third, Bayesian models inherently incorporate uncertainty, reflected in the posterior distributions and subsequently in the clustering outcomes. 
Having said that, we noted in our empirical analysis that the overall clustering structure was pretty stable across multiple chains. For critical applications, we recommend implementing a consensus clustering approach by aggregating results across multiple chains to identify the most robust and stable groupings.

A fully Bayesian approach was used for model estimation with a Metropolis-Hastings-within-Gibbs sampler. The \CRANpkg{lsirm12pl} package offers default estimation settings for priors, jumping rules, number of iterations, burn-ins, and thinning. The default estimation setting works reasonably well in a broad range of situations, but users can manually revise the estimation settings if desired. In addition, the package \CRANpkg{lsirm12pl} offers convenient supplemental functions to evaluate, summarize, visualize, diagnose, and interpret the estimated results. We provided detailed illustrations of the package with real data examples available in the package. We hope that these illustrations guide researchers in using the \CRANpkg{lsirm12pl} package for the analysis of their own datasets. 
%revealing their latent traits. A key feature is its interaction map, which visualizes these relationships in a low-dimensional space. 

The R package \CRANpkg{lsirm12pl} is our first step in making the LSIRM approach more applicable and usable in practice. The code is written using \CRANpkg{Rcpp} \citep{Dirk2011, Dirk2013, Dirk2018} and \CRANpkg{RcppArmadillo} \citep{Dirk2014} in R for efficient computation. For example, in our first data example with 726 respondents and 56 test items, the computation time for one chain using a single core \code{lsirm1pl} was 1.93 minutes on an Apple M1 laptop. We provide additional details of our package implementation in the package GitHub site (\href{https://github.com/jiniuslab/lsirm12pl}{https://github.com/jiniuslab/lsirm12pl}). We will continue to update the package by incorporating additional modeling, data analysis, and visualization options to make the \CRANpkg{lsirm12pl} package more useful in a wider range of situations. For example, we are currently engaged in the development of other extensions, such as for ordinal and longitudinal data. As advancements are made, we will systematically include them in the software package. This ongoing development will further enhance the utility and applicability of the LSIRM in various practical scenarios.
%Furthermore, %models for ordinal and longitudinal data are capable of expansion, and 
%the computation time was 2.06 min for the first data example with 726 respondents and 56 test items with \code{lsirm1pl}.

% % \section{Section title in sentence case}

% % Introductory section which may include references in parentheses
% % \citep{R}, or cite a reference such as \citet{R} in the text.

% % \section{Another section}

% % This section may contain a figure such as Figure~\ref{figure:rlogo}.

% % \begin{figure}[htb]
% %   \centering
% %   \includegraphics{Rlogo-5}
% %   \caption{The logo of R.}
% %   \label{figure:rlogo}
% % \end{figure}

% % \section{Another section}

% % There will likely be several sections, perhaps including code snippets, such as:

% % \begin{example}
% %   x <- 1:10
% %   result <- myFunction(x)
% % \end{example}

% % \section{Summary}

% % This file is only a basic article template. For full details of \emph{The R Journal} style and information on how to prepare your article for submission, see the \href{https://journal.r-project.org/share/author-guide.pdf}{Instructions for Authors}.

\section{Acknowledgement}

We thank the Editor, Associate Editor, and reviewers for their constructive comments. We also thank Dr. Won Chang for constructive comments on our work. This study was partially supported by the Basic Science Research Program through the National Research Foundation of Korea (NRF 2020R1A2C1A01009881, RS-2023-00217705, and RS-2024-00333701). Correspondence should be addressed to Ick Hoon Jin, Department of Applied Statistics, Department of Statistics and Data Science, Yonsei University, Seoul, Republic of Korea. Go, Kim, and Park are co-first authors and listed as an alphabetical order.

\bibliography{lsirm12pl}

\begin{thebibliography}{48}
\providecommand{\natexlab}[1]{#1}
\providecommand{\url}[1]{\texttt{#1}}
\expandafter\ifx\csname urlstyle\endcsname\relax
  \providecommand{\doi}[1]{doi: #1}\else
  \providecommand{\doi}{doi: \begingroup \urlstyle{rm}\Url}\fi

\bibitem[An and Yung(2014)]{An2014}
X.~An and Y.~F. Yung.
\newblock Item response theory: {W}hat it is and how you can use the {IRT}
  procedure to apply it.
\newblock \emph{SAS Institute Inc}, 10\penalty0 (4), 2014.

\bibitem[Andersen(1973)]{Andersen1973}
E.~B. Andersen.
\newblock {CONDITIONAL} {INFERENCE} {FOR} {MULTIPLE}-{CHOICE} {QUESTIONNAIRES}.
\newblock \emph{British Journal of Mathematical and Statistical Psychology},
  26\penalty0 (1):\penalty0 31--44, 1973.
\newblock \doi{10.1111/j.2044-8317.1973.tb00504.x}.

\bibitem[Andrich(1978)]{Andrich1978}
D.~Andrich.
\newblock A rating formulation for ordered response categories.
\newblock \emph{Psychometrika}, 43\penalty0 (4):\penalty0 561--573, 1978.
\newblock \doi{10.1007/bf02293814}.

\bibitem[Baker and Kim(2023)]{Baker2023}
F.~B. Baker and S.-H. Kim, editors.
\newblock \emph{Item response theory}.
\newblock Statistics: A Series of Textbooks and Monographs. Taylor \& Francis,
  London, England, 2 edition, 2023.

\bibitem[Batool and Hennig(2021)]{Batool2021}
F.~Batool and C.~Hennig.
\newblock Clustering with the {A}verage {S}ilhouette {W}idth.
\newblock \emph{Computational Statistics \& Data Analysis}, 158:\penalty0
  107190, 2021.
\newblock \doi{10.1016/j.csda.2021.107190}.

\bibitem[Bernaards and Jennrich(2005)]{Bernaards2005}
C.~A. Bernaards and R.~I. Jennrich.
\newblock Gradient {P}rojection {A}lgorithms and {S}oftware for {A}rbitrary
  {R}otation {C}riteria in {F}actor {A}nalysis.
\newblock \emph{Educational and Psychological Measurement}, 65\penalty0
  (5):\penalty0 676--696, Oct. 2005.
\newblock \doi{10.1177/0013164404272507}.

\bibitem[Birnbaum(1968)]{birnbaum1968}
A.~L. Birnbaum.
\newblock Some latent trait models and their use in inferring an examinee's
  ability.
\newblock \emph{Statistical theories of mental test scores}, 1968.

\bibitem[Braeken(2010)]{Braeken2010}
J.~Braeken.
\newblock A {B}oundary {M}ixture {A}pproach to {V}iolations of {C}onditional
  {I}ndependence.
\newblock \emph{Psychometrika}, 76\penalty0 (1):\penalty0 57–76, 2010.
\newblock ISSN 1860-0980.
\newblock \doi{10.1007/s11336-010-9190-4}.
\newblock URL \url{http://dx.doi.org/10.1007/s11336-010-9190-4}.

\bibitem[Brzezi\'{n}ska(2018)]{Brzezinska2018}
J.~Brzezi\'{n}ska.
\newblock Item {R}esponse {T}heory {M}odels in the {M}easurement {T}heory with
  the {U}se of ltm {P}ackage in {R}.
\newblock \emph{Econometrics}, 22\penalty0 (1):\penalty0 11--25, 2018.
\newblock \doi{doi:10.15611/eada.2018.1.01}.

\bibitem[Chalmers(2012)]{Chalmers2012}
R.~P. Chalmers.
\newblock mirt: {A} {M}ultidimensional {I}tem {R}esponse {T}heory package for
  the {R} {E}nvironment.
\newblock \emph{Journal of Statistical Software}, 48\penalty0 (6):\penalty0
  1–29, 2012.
\newblock \doi{10.18637/jss.v048.i06}.

\bibitem[Chen and Wang(2007)]{Chen2007}
C.-T. Chen and W.-C. Wang.
\newblock Effects of {I}gnoring {I}tem {I}nteraction on {I}tem {P}arameter
  {E}stimation and {D}etection of {I}nteracting {I}tems.
\newblock \emph{Applied Psychological Measurement}, 31\penalty0 (5):\penalty0
  388–411, 2007.
\newblock ISSN 1552-3497.
\newblock \doi{10.1177/0146621606297309}.
\newblock URL \url{http://dx.doi.org/10.1177/0146621606297309}.

\bibitem[Chib and Greenberg(1995)]{chib1995understanding}
S.~Chib and E.~Greenberg.
\newblock Understanding the metropolis-hastings algorithm.
\newblock \emph{{T}he {A}merican {S}tatistician}, 49:\penalty0 327--335, 1995.
\newblock \doi{https://doi.org/10.2307/2684568}.

\bibitem[de~Ayala(2009)]{Ayala2009}
R.~J. de~Ayala.
\newblock \emph{The theory and practice of item response theory}.
\newblock Methodology in the Social Sciences. Guilford Publications, New York,
  NY, 2009.

\bibitem[Eddelbuettel(2013)]{Dirk2013}
D.~Eddelbuettel.
\newblock \emph{Seamless {R} and {C++} {I}ntegration with {Rcpp}}.
\newblock Springer, New York, 2013.
\newblock ISBN 978-1-4614-6867-7.

\bibitem[Eddelbuettel and Balamuta(2018)]{Dirk2018}
D.~Eddelbuettel and J.~J. Balamuta.
\newblock Extending {R} with {C++}: {A} {B}rief {I}ntroduction to {Rcpp}.
\newblock \emph{The American Statistician}, 72\penalty0 (1):\penalty0 28--36,
  2018.
\newblock \doi{10.1080/00031305.2017.1375990}.

\bibitem[Eddelbuettel and Fran\c{c}ois(2011)]{Dirk2011}
D.~Eddelbuettel and R.~Fran\c{c}ois.
\newblock {Rcpp}: {S}eamless {R} and {C++} {I}ntegration.
\newblock \emph{Journal of Statistical Software}, 40\penalty0 (8):\penalty0
  1--18, 2011.
\newblock \doi{10.18637/jss.v040.i08}.

\bibitem[Eddelbuettel and Sanderson(2014)]{Dirk2014}
D.~Eddelbuettel and C.~Sanderson.
\newblock Rcpparmadillo: {A}ccelerating {R} with high-performance {C}++ linear
  algebra.
\newblock \emph{Computational Statistics and Data Analysis}, 71:\penalty0
  1054--1063, March 2014.
\newblock \doi{10.1016/j.csda.2013.02.005}.

\bibitem[Fischer and Parzer(1991)]{fischer1991}
G.~H. Fischer and P.~Parzer.
\newblock An extension of the rating scale model with an application to the
  measurement of change.
\newblock \emph{Psychometrika}, 56\penalty0 (4):\penalty0 637--651, 1991.
\newblock \doi{10.1007/bf02294496}.

\bibitem[Fischer and Ponocny(1994)]{fischer1994}
G.~H. Fischer and I.~Ponocny.
\newblock An extension of the partial credit model with an application to the
  measurement of change.
\newblock \emph{Psychometrika}, 59\penalty0 (2):\penalty0 177--192, 1994.
\newblock \doi{10.1007/bf02295182}.

\bibitem[Glas and Verhelst(1989)]{glas1989}
C.~A.~W. Glas and N.~D. Verhelst.
\newblock Extensions of the partial credit model.
\newblock \emph{Psychometrika}, 54\penalty0 (4):\penalty0 635--659, 1989.
\newblock \doi{10.1007/bf02296401}.

\bibitem[Goldberg(1992)]{Goldberg1992}
L.~R. Goldberg.
\newblock The development of markers for the {B}ig-{F}ive factor structure.
\newblock \emph{Psychological Assessment}, 4\penalty0 (1):\penalty0 26–42,
  1992.
\newblock ISSN 1040-3590.
\newblock \doi{10.1037/1040-3590.4.1.26}.
\newblock URL \url{http://dx.doi.org/10.1037/1040-3590.4.1.26}.

\bibitem[Golino(2016)]{TDRI}
H.~Golino.
\newblock Tdri dataset.csv, 2016.
\newblock URL
  \url{https://figshare.com/articles/dataset/TDRI_dataset_csv/3142321/1}.

\bibitem[Gower(1975)]{procruste}
J.~C. Gower.
\newblock Generalized procrustes analysis.
\newblock \emph{Psychometrik}, 40:\penalty0 33--51, 1975.
\newblock \doi{10.1007/bf02291478}.

\bibitem[Hoff et~al.(2002)Hoff, Raftery, and Handcock]{Hoff:2002}
P.~Hoff, A.~Raftery, and M.~S. Handcock.
\newblock Latent space approaches to social network analysis.
\newblock \emph{Journal of the American Statistical Association}, 97:\penalty0
  1090--1098, 2002.
\newblock \doi{10.1198/016214502388618906}.

\bibitem[Hohensinn(2018)]{Hohensinn2018}
C.~Hohensinn.
\newblock pc{IRT}: {A}n {R} {P}ackage for {P}olytomous and {C}ontinuous {R}asch
  {M}odels.
\newblock \emph{Journal of Statistical Software, Code Snippets}, 84\penalty0
  (2):\penalty0 1–14, 2018.
\newblock \doi{10.18637/jss.v084.c02}.

\bibitem[Ishwaran and Rao(2005)]{Ishwaran2005}
H.~Ishwaran and J.~S. Rao.
\newblock Spike and slab variable selection: Frequentist and {B}ayesian
  strategies.
\newblock \emph{The Annals of Statistics}, 33\penalty0 (2), Apr. 2005.
\newblock \doi{10.1214/009053604000001147}.
\newblock URL \url{https://doi.org/10.1214/009053604000001147}.

\bibitem[Jennrich(2002)]{Jennrich2002}
R.~I. Jennrich.
\newblock A simple general method for oblique rotation.
\newblock \emph{Psychometrika}, 67\penalty0 (1):\penalty0 7--19, Mar. 2002.
\newblock \doi{10.1007/bf02294706}.

\bibitem[Jeon et~al.(2021)Jeon, Jin, Schweinberger, and Baugh]{jeon2021mapping}
M.~Jeon, I.~H. Jin, M.~Schweinberger, and S.~Baugh.
\newblock Mapping unobserved item--respondent interactions: A latent space item
  response model with interaction map.
\newblock \emph{Psychometrika}, pages 1--26, 2021.
\newblock \doi{10.1007/s11336-021-09762-5}.

\bibitem[Karatzoglou et~al.(2004)Karatzoglou, Smola, Hornik, and
  Zeileis]{Karatzoglou2004}
A.~Karatzoglou, A.~Smola, K.~Hornik, and A.~Zeileis.
\newblock kernlab -- an {S4} package for kernel methods in {R}.
\newblock \emph{Journal of Statistical Software}, 11\penalty0 (9):\penalty0
  1--20, 2004.
\newblock \doi{10.18637/jss.v011.i09}.

\bibitem[Mair and Hatzinger(2007)]{Mair2007}
P.~Mair and R.~Hatzinger.
\newblock {E}xtended {R}asch {M}odeling: The e{R}m {P}ackage for the
  {A}pplication of {IRT} {M}odels in {R}.
\newblock \emph{Journal of Statistical Software}, 20\penalty0 (9):\penalty0
  1–20, 2007.
\newblock \doi{10.18637/jss.v020.i09}.

\bibitem[Masters(1982)]{masters1982}
G.~N. Masters.
\newblock A {R}asch model for partial credit scoring.
\newblock \emph{Psychometrika}, 47\penalty0 (2):\penalty0 149--174, 1982.
\newblock \doi{10.1007/bf02296272}.

\bibitem[McCullagh and Nelder(2019)]{mccullagh2019generalized}
P.~McCullagh and J.~A. Nelder.
\newblock \emph{Generalized linear models}.
\newblock Routledge, 2019.

\bibitem[M{\"u}ller(1987)]{Mller1987}
H.~M{\"u}ller.
\newblock A {R}asch model for continuous ratings.
\newblock \emph{Psychometrika}, 52\penalty0 (2):\penalty0 165--181, 1987.
\newblock \doi{10.1007/bf02294232}.

\bibitem[Myszkowski and Storme(2024)]{Myszkowski2024}
N.~Myszkowski and M.~Storme.
\newblock Modeling {S}equential {D}ependencies in {P}rogressive {M}atrices:
  {A}n {A}uto-{R}egressive {I}tem {R}esponse {T}heory ({AR-IRT}) {A}pproach.
\newblock \emph{Journal of Intelligence}, 12\penalty0 (1):\penalty0 7, 2024.
\newblock ISSN 2079-3200.
\newblock \doi{10.3390/jintelligence12010007}.
\newblock URL \url{http://dx.doi.org/10.3390/jintelligence12010007}.

\bibitem[Neyman and Scott(1952)]{Neyman1952}
J.~Neyman and E.~L. Scott.
\newblock A {T}heory of the {S}patial {D}istribution of {G}alaxies.
\newblock \emph{The Astrophysical Journal}, 116:\penalty0 144 --163, 1952.
\newblock \doi{10.1086/145599}.

\bibitem[Ng et~al.(2001)Ng, Jordan, and Weiss]{Ng2001}
A.~Y. Ng, M.~I. Jordan, and Y.~Weiss.
\newblock On {S}pectral {C}lustering: {A}nalysis and an {A}lgorithm.
\newblock In \emph{Proceedings of the 14th International Conference on Neural
  Information Processing Systems: Natural and Synthetic}, NIPS'01, page
  849–856, Cambridge, MA, USA, 2001. MIT Press.

\bibitem[Rasch(1960)]{rasch:60}
G.~Rasch.
\newblock \emph{Probabilistic models for some intelligence and attainment
  tests}.
\newblock Danish Institute for Educational Research, Copenhagen, 1960.

\bibitem[Rasch(1961)]{rasch1961}
G.~Rasch.
\newblock On general laws and the meaning of measurement in psychology.
\newblock In \emph{Proceedings of the {F}ourth {B}erkeley {S}ymposium on
  {M}athematical {S}tatistics and {P}robability, {V}olume 4: {C}ontributions to
  {B}iology and {P}roblems of {M}edicine}, pages 321--333, Berkeley, Calif.,
  1961. University of California Press.
\newblock URL \url{https://projecteuclid.org/euclid.bsmsp/1200512895}.

\bibitem[Rizopoulos(2006)]{Rizopoulos2006}
D.~Rizopoulos.
\newblock ltm: An {R} package for {L}atent {V}ariable {M}odeling and {I}tem
  {R}esponse {A}nalysis.
\newblock \emph{Journal of Statistical Software}, 17\penalty0 (5):\penalty0
  1–25, 2006.
\newblock \doi{10.18637/jss.v017.i05}.

\bibitem[Rubin(1976)]{rubin1976inference}
D.~B. Rubin.
\newblock Inference and missing data.
\newblock \emph{Biometrika}, 63\penalty0 (3):\penalty0 581--592, 1976.

\bibitem[Samejima(1968)]{Samejima1968}
F.~Samejima.
\newblock {ESTIMATION} {OF} {LATENT} {ABILITY} {USING} {A} {RESPONSE} {PATTERN}
  {OF} {GRADED} {SCORES1}.
\newblock \emph{ETS Research Bulletin Series}, 1968\penalty0 (1):\penalty0
  i--169, 1968.
\newblock \doi{10.1002/j.2333-8504.1968.tb00153.x}.

\bibitem[Scheiblechner(1972)]{scheiblechner1972}
H.~Scheiblechner.
\newblock Das {L}ernen und {L}{\"o}sen komplexer {D}enkaufgaben.
\newblock \emph{Zeitschrift f{\"u}r experimentelle und angewandte Psychologie},
  19:\penalty0 476--506, 1972.

\bibitem[Sewell and Chen(2015)]{Sewell:2015}
D.~K. Sewell and Y.~Chen.
\newblock Latent space models for dynamic networks.
\newblock \emph{Journal of the American Statistical Association}, 110:\penalty0
  1646--1657, 2015.

\bibitem[Tanner and Wong(1987)]{Tanner:1987p528}
M.~Tanner and W.~Wong.
\newblock The calculation of posterior distributions by data augmentation (with
  discussion).
\newblock \emph{Journal of the American Statistical Association}, 82:\penalty0
  528–550, 1987.

\bibitem[Thomas(1949)]{Thomas1949}
M.~Thomas.
\newblock A {G}eneralization of {P}oisson{\textquotesingle}s {B}inomial {L}imit
  {F}or use in {E}cology.
\newblock \emph{Biometrika}, 36\penalty0 (1/2):\penalty0 18--25, 1949.
\newblock \doi{10.2307/2332526}.

\bibitem[von Luxburg(2007)]{vonLuxburg2007}
U.~von Luxburg.
\newblock A tutorial on spectral clustering.
\newblock \emph{Statistics and Computing}, 17\penalty0 (4):\penalty0 395--416,
  2007.
\newblock \doi{10.1007/s11222-007-9033-z}.

\bibitem[Yi et~al.(2024)Yi, Kim, Park, Jeon, and Jin]{yi2023}
S.~Yi, M.~Kim, J.~Park, M.~Jeon, and I.~H. Jin.
\newblock Impacts of {I}nnovation {S}chool {S}ystem in {K}orea: {A} {L}atent
  {S}pace {I}tem {R}esponse {M}odel with {N}eyman-{S}cott {P}oint {P}rocess,
  2024.
\newblock URL \url{https://arxiv.org/abs/2306.02106}.

\bibitem[Zanon et~al.(2016)Zanon, Hutz, Yoo, and Hambleton]{Zanon2016}
C.~Zanon, C.~S. Hutz, H.~Yoo, and R.~K. Hambleton.
\newblock An application of item response theory to psychological test
  development.
\newblock \emph{Psicologia: Reflex{\~{a}}o e Cr{\'{\i}}tica}, 29\penalty0 (1),
  2016.
\newblock \doi{10.1186/s41155-016-0040-x}.

\end{thebibliography}

\address{Dongyoung Go\\
  Department of Statistics and Data Science\\
  Department of Applied Statistics\\
  Yonsei University\\
  50 Yonsei-ro, Seodaemun, Seoul, Republic of Korea\\
  \email{dongyoung.gr@yonsei.ac.kr}}

\address{Jina Park\\
  Department of Statistics and Data Science\\
  Department of Applied Statistics\\
  Yonsei University\\
  50 Yonsei-ro, Seodaemun, Seoul, Republic of Korea\\
  \email{pja0707@yonesi.ac.kr}}

\address{Gwanghee Kim\\
  Department of Statistics and Data Science\\
  Yonsei University\\
  50 Yonsei-ro, Seodaemun, Seoul, Republic of Korea\\
  \email{musagh08@yonesi.ac.kr}}

\address{Junyong Park\\
  Samsung Electronics, Formerly at Yonsei University\\
  1 Samsung Electronics-ro, Hwaseong-si, Gyeonggi-do, Republic of Korea\\
  \email{jun94.park@samsung.com}}

\address{Minjeong Jeon\\
  School of Education and Information Studies\\
  University of California, Los Angeles\\
  405 Hilgard Avenue, Los Angeles, CA 90095\\
  \email{mjjeon@ucla.edu}}

\address{Ick Hoon Jin\\
  Department of Statistics and Data Science\\
  Department of Applied Statistics\\
  Yonsei University\\
  50 Yonsei-ro, Seodaemun, Seoul, Republic of Korea\\
  \email{ijin@yonsei.ac.kr}}

\end{article}

\end{document}